\documentclass{article}
\usepackage{graphicx, bm, bbm,amsmath, amssymb, epsfig,color}
\usepackage[margin=1in]{geometry}
\pdfoutput=1

\newcommand{\nn}{\noindent}

\newcommand{\bq}{\begin{align}}
\newcommand{\eq}{\end{align}}

\title{Optimal wall shapes and flows for steady planar convection}
\author{Silas Alben}

\begin{document}

\maketitle

\begin{abstract}
We compute steady planar incompressible flows and wall shapes that maximize the rate of heat transfer (Nu) between and hot and cold walls, for a given rate of viscous dissipation by the flow (Pe$^2$). In the case of no flow, we show theoretically that the optimal walls are flat and horizontal, at the minimum separation distance. We use a decoupled approximation to show that flat walls remain optimal up to a critical nonzero flow magnitude. Beyond this value, our computed optimal flows and wall shapes converge to a set of forms that are invariant except for a Pe$^{-1/3}$ scaling of horizontal lengths. The corresponding rate of heat transfer Nu $\sim$ Pe$^{2/3}$. We show that these scalings result from flows at the interface between the diffusion-dominated and convection-dominated regimes. We also show that the separation distance of the walls remains at its minimum value at large Pe. 
\end{abstract}

\section{Introduction}

The transfer of heat from solid boundaries to adjacent fluid flows is fundamental to many natural and technological processes. Important examples include the formation and evolution of stars and planets, and energy consumption in buildings, computer heat sinks, and organisms \cite{blundell2010concepts,bergman2011fundamentals}. For flows with large spatial scales and/or large flow speeds, the heat transfer is often controlled by thin viscous and thermal layers at the solid boundaries. Consequently, the rate of heat transfer is sensitive to the features of the solid boundaries, and in particular their shape or geometry \cite{webb2005enhanced,lienhard2013heat,wen2022steady,wen2022heat,tobasco2022optimal,song2023bounds}. 

The effect of wall shape modulations or roughness on thermal transport has been studied many times both for natural and forced convection.
Toppaladoddi {\it et al.} considered natural convection in a fluid layer between hot and cold sinusoidal surfaces \cite{toppaladoddi2017roughness}. 
They fixed the roughness amplitude at one-tenth the fluid layer height, and found that heat transfer was maximized when the roughness amplitude and wavelength were similar. This is one example of a large body of numerical and experimental work on the effect of wall roughness on the rate of heat transfer in turbulent natural convection with various roughness geometries---sinusoidal \cite{toppaladoddi2015tailoring,toppaladoddi2017roughness,zhu2017roughness}, triangular \cite{zhang2018surface}, cubic \cite{rusaouen2018thermal}, ratchet-shaped \cite{jiang2018controlling}, ring-shaped \cite{emran2020natural}, and fractal \cite{toppaladoddi2021thermal} and other multiscale roughness profiles \cite{zhu2019scaling,sharma2022investigation}. A main focus of these works is the asymptotic scaling of the rate of heat transfer with the Rayleigh number and its dependence on the geometric form of the roughness and its length scales including wavelengths and heights of the roughness profiles \cite{yang2021dependence}. An experimental work that considered pyramid-shaped roughness elements of varying aspect ratio related the heat transfer enhancement to the dynamics of thermal plumes near the roughness elements \cite{xie2017turbulent}.

Variations of these problems include the effect of tilting the rough walls, for triangular \cite{chand2022effect} and ratched-shaped surfaces \cite{jiang2023effects}, and the heat transfer enhancement that can be obtained by moving the rough plates \cite{jin2022shear}. \cite{zhang2018surface} showed computationally that in some cases with small roughness heights, roughness can actually decrease the rate of heat transfer. On the theoretical side, Goluskin and Doering used the ``background method" to obtain upper bounds on the rate of heat transfer in fluid layers between upper and lower rough walls whose profiles correspond to single-valued functions of the horizontal coordinate \cite{goluskin2016bounds}. Another theoretical work derived upper bounds on heat transfer for Navier-slip rough boundaries \cite{bleitner2022bounds}.

A related body of work has relaxed the requirement that the flow solve the Boussinesq equations for natural convection, and instead optimized the rate of heat transfer over the larger class of all incompressible flows between hot and cold boundaries \cite{hassanzadeh2014wall,souza2016optimal,tobasco2017optimal,marcotte2018optimal,motoki2018maximal,motoki2018optimal,doering2019optimal,souza2020wall,kumar2022three,alben2023transition}. These works consider the simple geometry---inspired by Rayleigh-B\'{e}nard convection---consisting of a layer of fluid between flat horizontal walls at different fixed temperatures. The optimal flows have also been calculated for domains obtained by conformal mappings \cite{alben2017optimal} and in flows through channels \cite{alben2017improved}. All incompressible flows are solutions to the incompressible Navier-Stokes equations with a certain distribution of force per unit volume applied over the flow domain \cite{alben2017improved}. Although the forcing distribution may be difficult to create exactly in an experiment, simple approximate distributions may be sufficient to obtain large improvements from previous flows \cite{alben2017improved}. The flows can be used to identify typical features of efficient flows, such as branching structures \cite{zimparov2006thermodynamic,tobasco2017optimal,motoki2018maximal,alben2023transition}.

A large body of engineering work has studied the effect of wall roughness in simple forced convection scenarios. The geometry of walls separating two fluids in a heat exchanger is often manipulated to increase the rate of heat transfer for a given amount of power needed to drive the flow \cite{webb2005enhanced,bergles2013current}. A typical strategy is the insertion of helical coils in tubes with axial flow \cite{gee1980forced}. Another well-studied example is the enhancement of heat transfer due to sinusoidal wall modulations in channel flow, and the fluid-dynamical mechanisms underlying the enhancement
\cite{Castelloes2010,Sui2010, Gong2011,Kanaris2006,Ramgadia2013,Guzman2009,Muthuraj2010,
Stone2004,grant2014onset}. 


In this work we again consider optimal incompressible flows for heat transfer between hot and cold walls \cite{hassanzadeh2014wall,souza2016optimal,tobasco2017optimal,motoki2018maximal,souza2020wall,kumar2022three,alben2023transition}, but now in the presence of wall roughness. In particular, we extend the computational approach of \cite{alben2023transition} to search for optimal flows together with optimal boundary wall shapes, in order to maximize the rate of heat transfer (Nu) given a certain rate of power consumption by the flow (Pe$^2$). At zero power consumption, flat walls are shown to be optimal theoretically in section \ref{sec:ZeroPe}. Below a certain power consumption rate, the computed optimal flows are rectangular convection rolls with flat walls, as in previous studies \cite{souza2020wall, alben2023transition}. In section \ref{sec:SmallPe} this is shown theoretically by showing that the leading-order effects of the flows and the roughness are decoupled when both are small. 

In section \ref{sec:Computations} we describe the numerical methods for computing temperature fields and optimal flows. Using these methods, in section \ref{sec:Resolution} we show the solution features---sharp temperature gradients---that limit the accuracy of the computations, and how the accuracy depends on key wall shape parameters. We also use the computations to test the accuracy of the decoupled approximation at small Pe and wall deflection amplitudes, in section \ref{sec:TestDecoupled}.

We present computed optima at moderate and large Pe in section \ref{sec:LargePe}. Above a critical Pe, the computed optima with wavy walls outperform those with flat walls. At large power consumption rates, the optimal flow streamlines and wall shapes fall within a large but well-defined set of typical configurations. The configurations are invariant at large Pe, except for a power-law scaling of their horizontal period together with an O(1) vertical roughness of the walls. Thus the convection rolls are very elongated vertically in the limit of large Pe. The scaling of the horizontal period, $L_x \sim$ Pe$^{-1/3}$, is close to that seen for the flat-wall optima in previous work, and is shown here to result in flows at the interface between diffusion-dominated and convection-dominated regimes. The corresponding heat transfer rate Nu scales as Pe$^{2/3}$, which was shown to be an upper bound in the flat-wall case \cite{souza2016optimal} in 2D and 3D flows, and has been observed in 3D flows with flat walls \cite{motoki2018maximal} but only up to logarithmic corrections in 2D flows with flat walls \cite{tobasco2017optimal}. The Pe$^{2/3}$ scaling corresponds to an upper bound for flows with rough walls shown theoretically by \cite{goluskin2016bounds}.

In section \ref{sec:MinH} we explain why the optima have the minimum possible separation between the hot and cold walls, and section \ref{sec:Conclusions} presents the conclusions and additional context for the results.


\section{Model \label{sec:Model}}

\begin{figure}[h]
    \centering
    \includegraphics[width=4in]{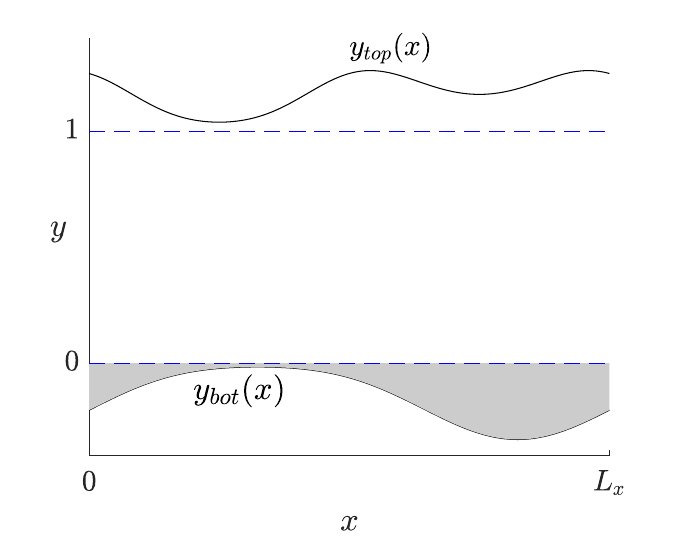}
    \caption{\footnotesize Example of a domain that is periodic in $x$ with period $L_x$. The bottom boundary is $y_{bot}(x) \leq 0$ and the top boundary is $y_{top}(x) \geq 1$. The boundaries do not cross the blue dashed lines. The gray region is indicated for the discussion in section \ref{Pe0}.}
    \label{fig:PureConductionSchematicFig}
             \vspace{-.10in}
\end{figure}

A 2D layer of fluid is contained in the region $-\infty < x < \infty$ and $y_{bot}(x) < y < y_{top}(x)$ (see figure \ref{fig:PureConductionSchematicFig}). The curvilinear walls at $y = y_{bot}(x)$ and $y = y_{top}(x)$ have temperatures 1 and 0 respectively. The two walls are periodic in $x$ with a common period $L_x$ that will be determined during the optimization.  The walls (black lines in figure \ref{fig:PureConductionSchematicFig}) are constrained so $y_{bot} \leq 0$ and $y_{top} \geq 1$  respectively; i.e. the walls may not cross the blue dashed lines. Without this constraint, the distance between the walls could approach zero and then the temperature gradient would diverge, generally along with the rate of heat transfer, regardless of the fluid flow. The constraint allows us to ask which aspects of the walls' shapes other than their separation distance are useful for heat transfer by the intervening fluid. One can also interpret the constraint as choosing a characteristic length scale for the problem, which is $H_{min}$, the minimum allowed separation distance between the walls. The separation distance $H \equiv \min y_{top}(x) - \max y_{bot}(x)$ is constrained to be$ \geq H_{min}$, or 1 when we nondimensionalize all distances by $H_{min}$, going forward. Although $H > 1$ is possible, we find that $H = 1$ for all computed optima. Intuitively, minimizing the separation distance between the walls seems advantageous for heat transfer, and in section \ref{sec:MinH} we use a scaling argument to explain why this choice might be optimal. Also, the choice of wall temperatures (1 and 0) is equivalent to setting the characteristic temperature scale to be the difference between the wall temperatures. In addition to the constraint $H \geq 1$, the walls' vertical positions must be single-valued functions of $x$ but are otherwise essentially arbitrary.

Between the walls, the temperature field is determined by the steady advection-diffusion equation 
\begin{equation}
    \mathbf{u}\cdot\nabla T - \nabla^2T = 0. \label{AdvDiff}
\end{equation}
driven by an incompressible flow field
$(u,v) = \mathbf{u} = (\partial_y\psi,-\partial_x\psi)$, with $\psi(x,y)$ the stream function. We have set the prefactor of $\nabla^2T$ (the thermal diffusivity) to unity, corresponding to nondimensionalizing velocities by $\kappa/H_{min}$, with $\kappa$ the dimensional thermal diffusivity.

The setup is similar to Rayleigh-B\'{e}nard convection, but instead of solving the Boussinesq equations for the flow field, we consider all incompressible flows and find those that maximize the Nusselt number, the heat transfer from the hot lower boundary (per unit horizontal distance), 
\begin{align}
    \mbox{Nu} = \frac{1}{L_x}\int_{C = \{(x, y_{bot}(x))\}} \partial_n T ds . \label{Nu}
\end{align}
Integrating $\nabla^2 T$ over the whole region and using periodicity at the side walls, we see that Nu is also the heat transfer into the upper (cold) boundary.

We maximize Nu over flows with a given spatially-averaged rate of viscous dissipation Pe$^2$:
\begin{equation}
    \mbox{Pe}^2 \equiv \frac{1}{2}\frac{1}{L_x}\int_0^{L_x}\int_{y_{bot}(x)}^{y_{top}(x)} \left(\nabla \mathbf{u} + \nabla \mathbf{u}^T\right)^2 dy \, dx = \frac{1}{L_x}\int_0^{L_x} \int_{y_{bot}(x)}^{y_{top}(x)}  \left(\nabla^2 \psi\right)^2 dy \, dx. \label{Power}
\end{equation}
Pe$^2$ is the average rate of viscous dissipation per unit horizontal width, nondimensionalized by $\mu \kappa^2/H_{min}^3$, with $\mu$ the viscosity.
The second equality in (\ref{Power}) holds for incompressible flows given the no-slip boundary conditions at the top and bottom walls and periodicity in $x$ we have here \cite{lamb1932hydrodynamics,Mil68}. 

An incompressible flow field that maximizes Nu for a given Pe can also be considered a solution to the Boussinesq or Navier-Stokes equations with a suitable forcing function---whatever forcing is needed to balance the remaining terms in the equations. 
The optimization problem here is essentially the same as in \cite{souza2020wall,alben2023transition} except that the walls are no longer flat and horizontal but must be determined together with the flow field and the horizontal period. 

Like the walls, the flow $\psi$ is periodic in $x$ with period $L_x$. In terms of $\psi$, the no-slip boundary conditions are: $\psi = \partial_n\psi = 0$ at $y = y_{bot}(x)$ and $\psi = \psi_{top}$ and $\partial_n\psi = 0$ at $y = y_{top}(x)$. The constant $\psi_{top}$ is the net horizontal fluid flux through the channel. In most of our computed optima including those with the highest Nu at a given Pe, $\psi_{top}$ is very close to 0. 


\section{No convection (Pe = 0) \label{Pe0} \label{sec:ZeroPe}}

First we consider the limiting case Pe = 0, the pure-conduction problem. We show that the maximum Nu is achieved with flat walls with the minimum separation, $y_{bot}(x) \equiv 0$ and $y_{top}(x) \equiv 1$. Using $\nabla^2 T = 0$, we show that the net vertical conductive heat flux across all horizontal cross-sections is constant over the portion of $y$ values that the walls don't cross, $0 \leq y \leq 1$:
\begin{align}
     \frac{d}{dy}\int_0^{L_x} \partial_y T(x,y)\, dx = \int_0^{L_x} \partial_{yy} T(x,y) \,dx =  \int_0^{L_x} -\partial_{xx} T(x,y) \,dx = 0, \, 0 \leq y \leq 1
\end{align}
using the periodicity in $x$ for the last equality. Thus
\begin{align}
     \int_0^{L_x} &\partial_y T dx = \{ \mbox{constant in } y:  0 \leq y \leq 1 \} = \int_0^1 \int_0^{L_x}  \partial_y T dx \, dy \label{flux1} \\ & =  \int_0^{L_x}  \int_0^1 \partial_y T  dy \, dx  = \int_0^{L_x} T(x,1) - T(x,0)\, dx. \label{flux2}
\end{align}
If at some point $x_0$ $y_{bot}(x_0) < 0$ or $y_{top}(x_0) > 1$, then for some open $x$-interval containing $x_0$ we have $T(x,0) < 1$ or 
$T(x,1) > 0$ respectively, by the maximum principle for harmonic functions---the maximum and minimum of $T$ are taken on the boundary. In that case $T(x,0) - T(x,1)$, which is $\leq 1$ for any wall shape by the maximum principle, is in fact strictly less than 1 on some $x$-interval. Therefore its average (over $x$) is strictly less than 1:
\begin{align}
1 > \frac{1}{L_x}\int_0^{L_x} T(x,0) - T(x,1) \, dx = \frac{1}{L_x}\int_0^{L_x} -\partial_y T|_{y = 0} dx. \label{Tdiff}  
\end{align}
The last equality in (\ref{Tdiff}) follows by (\ref{flux1})--(\ref{flux2}).
We can show that the last quantity in (\ref{Tdiff}) is in fact Nu, by again integrating $\nabla^2 T$, this time over the gray region in figure \ref{fig:PureConductionSchematicFig}. We have
\begin{align}
    0 &= \frac{1}{L_x}\iint \nabla^2 T dA = \frac{1}{L_x} \oint \partial_n T ds \label{gray1} \\0 &= \frac{1}{L_x}\int_{C = \{(x, y_{bot}(x))\}} \partial_n T ds + \frac{1}{L_x} \int_0^{L_x} \partial_y T|_{y = 0} dx \label{gray2}
\end{align}
The contributions to the closed contour integral in (\ref{gray1}) from the vertical sides of the gray region cancel by $x$ periodicity. Combining (\ref{Nu}), (\ref{Tdiff}), and (\ref{gray2}), we have Nu $<$ 1 for nonflat walls, whereas Nu = 1 for flat walls (in which case $-\partial_y T \equiv 1$), the optimal solution with no flow (Pe = 0).

\section{Small convection ($0 <$ Pe $\ll 1$) \label{sec:SmallPe}}

Now we consider the case $0 <$ Pe $\ll 1$. We denote the deviations of the bottom and top walls from their flat states by $H_1$ and $H_2$:
\begin{align}
H_1(x) \equiv -y_{bot} \geq 0 \; ; \; H_2(x) \equiv y_{top}(x) - 1 \geq 0.    
\end{align}
For this asymptotic analysis and for the subsequent computations, we define $(p,q)$ coordinates, in which the flow domain is fixed as a unit square, for all wall shapes and all $L_x$:
\begin{align}
p \equiv \frac{x}{L_x} \; ; \; q \equiv \frac{y - y_{bot}(x)}{y_{top}(x) - y_{bot}(x)} = \frac{y + H_1(x)}{1+H_1(x) + H_2(x)}. \label{pq}
\end{align}
In the limit Pe $\to 0$, the optimal combination of flow and wall shapes must have $H_1, H_2 = o(1)$, i.e. tend to case of flat walls, by the argument in the previous section. Otherwise Nu would be strictly less than 1 in the limit, whereas the flat-wall case has Nu $\geq 1$ for any incompressible flow field \cite{tobasco2017optimal}.

When we write the advection-diffusion equation (\ref{AdvDiff}) in $(p,q)$ coordinates, the small perturbation to flat walls corresponds to adding small space-varying coefficients in front of the differential operators. At small Pe the flow also takes the form of a small space-varying coefficient in (\ref{AdvDiff}). We write 
\begin{align}
    \mathbf{u} = \mathbf{u}_1 + O\left(\mbox{Pe}^2\right) \, ; \, \nabla_{x,y} = \nabla_{p,q} + \nabla_1 + ... \, ; \, \nabla_{x,y}^2 = \nabla_{p,q}^2 + \nabla^2_1 + ... 
\end{align}
where $\mathbf{u}_1 = O(\mbox{Pe})$ and $\nabla_1$ and $\nabla^2_1$ are first and second-order differential operators in $p$ and $q$ with coefficients that are linear in $H_1$ and $H_2$, and therefore o(1) as Pe $\to 0$. Terms with quadratic and higher-order coefficients are subdominant and not written explicitly. Given the flow and wall shape, the optimal temperature field has an expansion
\begin{align}
    T = T_0 + T_1 + \ldots 
\end{align}
with $T_0 = 1-q$ and $T_1 = o(1)$.
We insert the expansions of $\mathbf{u}$, $T$, and the differential operators into the advection-diffusion equation (\ref{AdvDiff}). At $O$(Pe$^0$),
\begin{align}
    -\nabla_{p,q}^2 T_0 = 0
\end{align}
so $T_0 = 1-q$, the flat-wall steady-conduction solution. We then take the next-order terms, linear in Pe or the wall-perturbation amplitude,
\begin{align}
    -\nabla_{p,q}^2 T_1 = \nabla_1^2 T_0 - \mathbf{u}_1 \cdot \nabla_{p,q} T_0.
\end{align}
Thus $T_1$ satisfies Poisson's equation forced by a superposition of the wall perturbation and the small flow. We decompose $T_1$ as the sum of the perturbations from each forcing term separately:
\begin{align}
    T_1 &= T_{1A} + T_{1B} \\
    -\nabla_1^2 T_{1A} &= \nabla_1^2 T_0 \\
    -\nabla_1^2 T_{1B} &= - \mathbf{u}_1 \cdot \nabla_{p,q} T_0.
\end{align}
$T_{1A}$ is the leading-order change in the temperature field with the wall perturbation and zero flow, while $T_{1B}$ is the leading-order change in the temperature field with the $O$(Pe) flow and flat walls.

The Nusselt number
\begin{align}
    \mbox{Nu} &= \frac{1}{L_x}\int_0^{L_x}\hat{\mathbf{n}}\cdot \nabla T \,\frac{ds}{dx} \bigg|_{y = y_{bot}(x)} dx= \int_0^{1} \hat{\mathbf{n}}\cdot \nabla T \,\frac{ds}{dx} \bigg|_{q=0} dp
\end{align}
is expanded similarly, using the expansion for $T$. The wall perturbation results in an expansion
 \begin{align}
\frac{ds}{dx}\,\hat{\mathbf{n}}\cdot \nabla |_{q=0} =
\partial_q + \left(\frac{ds}{dx}\,\hat{\mathbf{n}}\cdot \nabla\right)_1 + \ldots
 \end{align}
The subscript 1 again denotes the leading-order perturbation from the flat-wall term, $\partial_q$ in this case since $ds/dx = 1$ for the flat wall.
The leading-order terms in the expansion of Nu are
\begin{align}
    \mbox{Nu} &= \int_0^{1} \partial_q T_0 |_{q=0} dp +
    \int_0^{1} \left(\frac{ds}{dx}\,\hat{\mathbf{n}}\cdot \nabla\right)_1 T_0 \bigg|_{q=0} dp + 
    \int_0^{1} \partial_q T_{1A} dp + 
    \int_0^{1} \partial_q T_{1B} dp + \ldots. \label{NuExpn1}\\
    &= \mbox{Nu}_0 + \mbox{Nu}_{1A} + \mbox{Nu}_{1B} + \ldots. \label{NuExpn2}
\end{align}
The first integral in (\ref{NuExpn1}) is Nu$_0$, the purely conductive heat transfer with flat walls, and equals 1. The sum of the second and third integrals in (\ref{NuExpn1}) is Nu$_{1A}$, the leading-order correction due to nonflat walls, and the fourth integral in (\ref{NuExpn1}) is Nu$_{1B}$, the leading-order correction due to the flow. For small wall perturbations and small Pe, the leading-order correction due to nonflat walls, Nu$_{1A}$, is independent of the flow. It is the change in the pure conduction heat transfer due to nonflat walls. In the previous section (Pe = 0) it was shown that this change can only decrease Nu; i.e. Nu$_{1A} < 0$. 

Furthermore, the leading-order correction due to the flow, Nu$_{1B}$, is independent of the wall perturbation. It is the leading-order change in Nu due to the optimal flow with flat walls at small Pe, which is a periodic array of almost square convection rolls \cite{hassanzadeh2014wall,souza2016optimal,alben2023transition}. It turns out that even though $T_{1B} = O$(Pe), Nu$_{1B}$ is smaller, $O$(Pe$^2$) \cite{hassanzadeh2014wall,souza2020wall,alben2023transition}. However, any $o(1)$ form of Nu$_{1B}$ is consistent with the expansion (\ref{NuExpn2}), which gives the decoupled effect of the walls and flow at leading order. The other main ingredient for (\ref{NuExpn2}) was that the optimal wall perturbation amplitude is $o(1)$ as Pe $\to 0$, as shown previously.

The main consequence of 
(\ref{NuExpn2}) that we highlight is that nonflat walls can only be optimal when Pe is large enough for the remainder terms in (\ref{NuExpn2}) to be comparable in magnitude to Nu$_{1B}$, so they outweigh its negative effect on Nu. Therefore, the transition to optima with nonflat walls occurs only above some critical Pe $> 0$ and not at arbitrarily small Pe. 

In the next section, we describe our computational methods for solving for temperature fields, testing the accuracy of the decoupled approximation (\ref{NuExpn2}), and computing optimal flows and wall shapes across the transition to nonflat walls.

\section{Computational methods \label{sec:Computations}}

In order to solve (\ref{AdvDiff}) with a variety of wall shapes and horizontal periods, we use $(p,q)$ coordinates (\ref{pq})
so the computational domain is square. We write each of equations (\ref{AdvDiff})--(\ref{Power}) in $(p,q)$ coordinates, and the differential operators and integrals in $(x,y)$ are replaced with differential operators and integrals in $(p,q)$ multiplied by functions of $L_x$, $y_{bot}(p)$, $y_{top}(p)$, and $q$ and their first and second derivatives with respect to $p$ and $q$. 

We define the wall deformations to be single-signed and bounded using auxiliary functions $h_1$ and $h_2$:
\begin{align}
    y_{bot}(p) = -A \left(\frac{1+\sin{A_1}}{2}\right)\frac{h_1(p)^2}{\|h_1(p)^2\|_6} \; , \;
    y_{top}(p) = A \left(\frac{1+\sin{A_2}}{2}\right)\frac{h_2(p)^2}{\|h_2(p)^2\|_6} \label{WallShapes}
\end{align}
$A$ is approximately the maximum amplitude of $y_{bot}$ and $y_{top}$, approximate because $\|\cdot\|_6$ instead of $\|\cdot\|_\infty$ appears in the denominators. This is done to allow for a smooth dependence of $y_{bot}$ and $y_{top}$ on $h_1$ and $h_2$, which is needed to calculate first derivatives in our quasi-Newton optimization method. An approximate maximum amplitude is sufficient because the optima become insensitive to $A$ above a certain threshold, and we run the simulations for a range of large $A$. The parameters $A_1$ and $A_2$ allow the approximate amplitudes to take any value between 0 and the maximum, $A$.
$h_1(p)$ and $h_2(p)$ are defined as Fourier series in $p$ with wave numbers ranging from 0 to $M_1$, which results in $2M_1 + 1$ modes (sines and cosines) for each function. The Fourier coefficient vectors are denoted $\mathbf{B}_1$ and
$\mathbf{B}_2$ respectively. 

For $L_x$ we use the expression
\begin{align}
    L_x = \min\left(5.4 \mbox{Pe}^{-0.37},0.5\right)+1+\sin(L_0);
\end{align}
which, for arbitrary real $L_0$, constrains $L_x$ to be non-negative and bounded below by one of the terms in the min function. The first of these is chosen to be about half the value for the flat-wall case at large Pe from \cite{alben2023transition}, and the second is used at small Pe. The nonzero lower bound is used to avoid spurious optima with very inaccurate $\partial_n T$ that appear at very small $L_x$.  

We express the flow $\psi$ using the same modes as in \cite{alben2023transition} but in $(p,q)$ coordinates instead of $(p,y)$ coordinates. We also include modes that give a net flow through the channel for greater generality, though their coefficients turn out to be essentially zero in the optima, as was found in preliminary computations in \cite{alben2023transition}. For $\psi$, we first form $\tilde{\psi}$, a linear combination of functions 
with period $p$ that have zero first derivatives in $p$ and $q$ at the walls. The functions are products of Fourier modes in $p$ and
linear combinations of Chebyshev polynomials in $q$:
\begin{align}
     \tilde{\psi}(p,q) &= \tilde{A}(3q^2-2q^3) + \sum_{j = 0}^{M}\sum_{k = 1}^{N-3} A_{jk}Q_k(q)\cos(2\pi j p) + B_{jk}Q_k(q)\sin(2\pi j p) \label{Modes} \\
     &Q_k(q) \in \langle T_0(2q-1), \ldots, T_{k+4}(2q-1) \rangle. 
\end{align}
The first term, with coefficient $\tilde{A}$, gives a net flux through the channel, and the remaining terms modify the flow distribution without changing the net flux. $Q_k$ is a linear combination of Chebyshev polynomials of the first kind up to degree $k+4$ that have zero values and first derivatives at $q = 0$ and 1. Its computation is described in appendix B of \cite{alben2023transition}. The functions $\tilde{\psi}$, $y_{bot}$, and $y_{top}$ can be shifted by the same arbitrary amount in $p$ without changing the solution. To remove this degree of freedom we set $B_{11} = 0$. Since $\partial_p \tilde{\psi} = \partial_q \tilde{\psi} = 0$ at the walls, the first derivatives in the tangential and normal directions ($\partial_s \tilde{\psi}$, $\partial_n \tilde{\psi}$) are zero there as well, so no-slip conditions are obeyed.

We define a grid that is uniform in $p$ with $m$ points, $\{0, 1/m, ..., 1-1/m\}$. Typically $m$ = 256. We concentrate points near the boundaries in $q$, in case sharp boundary layers appear in the optimal flows as in \cite{alben2023transition}. This is done by starting with a uniform grid for $\eta \in [0, 1]$ with spacing $1/n$, and mapping to the $q$-grid by
\begin{equation}
q = \eta -\frac{q_f}{2\pi}\sin(2\pi\eta).   \label{qeta}
\end{equation}
with $q_f$ a scalar. The $q$-spacing is maximum, $\approx (1+q_f)/n$, near $q = 1/2$, and minimum, $\approx (1-q_f)/n$, near $q = 0$ and 1. We take $q_f = 0.997$ and
$n$ = 256--1024, giving a grid spacing $\Delta q \approx 3 \times 10^{-6}$--$1.2 \times 10^{-5}$ at the boundaries. The derivative operators are discretized with second-order finite differences on these grids.

We obtain $\psi$ from $\tilde{\psi}$ by normalizing it so the flow has the power dissipation rate Pe$^2$ in (\ref{Power}). We do this in discrete form,  discretizing the derivatives in (\ref{Power}) with second-order finite differences and the integrals with the trapezoidal rule in $(p,q)$.

The discretized $\psi$ is written $\mathbf{\Psi}$, a vector of values at the $m(n-1)$ interior mesh points for $(p,q) \in [0, 1) \times (0,1)$. To form $\mathbf{\Psi}$, we arrange the modes in (\ref{Modes}) as columns of a $m(n-1) \times (2M+1)(N-3)-1$ matrix $\mathbf{V}$. Here we take $M = 5m/32$ and $N = 5n/32$, so we limit the modes to those whose oscillations can be resolved by the mesh. We set
$\tilde{\mathbf{\Psi}} = \mathbf{V}\mathbf{c}$, a linear combination of the discretized modes with coefficients $\mathbf{c}$.  To form a $\mathbf{\Psi}$ with power Pe$^2$, we discretize the integral in (\ref{Power}) as a quadratic form $\mathbf{a}^T\mathbf{M}\mathbf{a}$, where the vector $\mathbf{a}$ stands for a discretized $\psi$ in (\ref{Power}), and the matrix $\mathbf{M}$ gives the effect of the discretized derivatives and integrals in (\ref{Power}). Then
\begin{equation}
\mathbf{\Psi} = \frac{\mbox{Pe}\, \mathbf{V}\mathbf{c}}{\sqrt{(\mathbf{V}\mathbf{c})^T\mathbf{M}\mathbf{V}\mathbf{c}}} \label{Psi}
\end{equation}
has power Pe$^2$, as can be seen by evaluating $\mathbf{\Psi}^T\mathbf{M}\mathbf{\Psi}$. Such a $\mathbf{\Psi}$ automatically gives an incompressible flow by the stream function definition, and automatically satisfies the power constraint.

The various constraints on the walls' shapes and the flows have been enforced implicitly in the chosen forms of the functions that describe them. We are left with an unconstrained optimization problem to find the values of the parameters or coefficients that appear in these functions, and which may take any real values. Thus we maximize Nu ((\ref{Nu}), discretized at second-order) over $\mathbf{c} \in \mathbb{R}^{(2M+1)(N-3)-1}$, $\{\mathbf{B_1},\mathbf{B_2}\} \in \mathbb{R}^{2M_1+1}$, and $\{A_1, A_2, L_0\} \in \mathbb{R}$. We compute optima over a range of Pe, and use various combinations for $M_1$ and $A$ (the maximum wall deformation amplitude, approximately). We find empirically that for a given $M_1$, if $A$ is too large, the algorithm converges to spurious optima that are underresolved. We study this phenomenon using model problems in section \ref{sec:Resolution}.

We solve the optimization problem using the BFGS method \cite{martins2021engineering}, a quasi-Newton method that requires evaluations of the objective function Nu and its gradient with respect to the design parameters,  $\{\mathbf{c}, \mathbf{B_1},\mathbf{B_2}, A_1, A_2, L_0\}$. Nu is computed by forming $\mathbf{\Psi}$ from the design parameters, then computing $\mathbf{u}$ and solving (\ref{AdvDiff}) for the discretized temperature field $\mathbf{T}$ using second-order finite differences. We use a second-order rather than spectral discretization to obtain a sparse matrix in the advection-diffusion equation, allowing for relatively fast solutions.

The gradient can be computed efficiently using the adjoint method \cite{martins2021engineering}. The procedure is the same as that described in
\cite{alben2023transition}, but now including the wall shape parameters. In appendix \ref{sec:Adjoint} we present formulas for the gradient of Nu with respect to the parameters using the adjoint variable.

We initialize with about 100 random sets of $\{\mathbf{c}, \mathbf{B_1},\mathbf{B_2}, A_1, A_2, L_0\}$ with $M_1 = 1-4$ at various $A$. We run the optimization until the 2-norm of the gradient of Nu is less than $10^{-10}$ or the number of iterations reaches 10,000. The latter case is more common at larger Pe, where the gradient is larger in the initial random state, and where convergence to very small gradients is more difficult. Here the gradient norm is often $\approx$ 10$^{-2}$ after 10,000 iterations. At large Pe, the initial gradient norm is typically between 10$^3$ and 10$^4$, so a gradient norm of 10$^{-2}$ corresponds to 10$^{-5}$--10$^{-6}$ relative to the initial norm. The discretized advection-diffusion equation becomes increasingly ill-conditioned as Pe increases, causing a decrease of accuracy in Nu and its gradient, and slowing convergence \cite{kelley1999iterative}. Nonetheless, we are able to obtain accurate Nu values up to Pe = $10^{7}$, the largest value used with nonflat walls in this study. It is possible to obtain accurate optima at larger Pe, but the large-Pe behavior is already clear at Pe = $10^{7}$.

\subsection{Boundary heat flux resolution \label{sec:Resolution}}

\begin{figure}[h]
    \centering
    \includegraphics[width=6.5in]{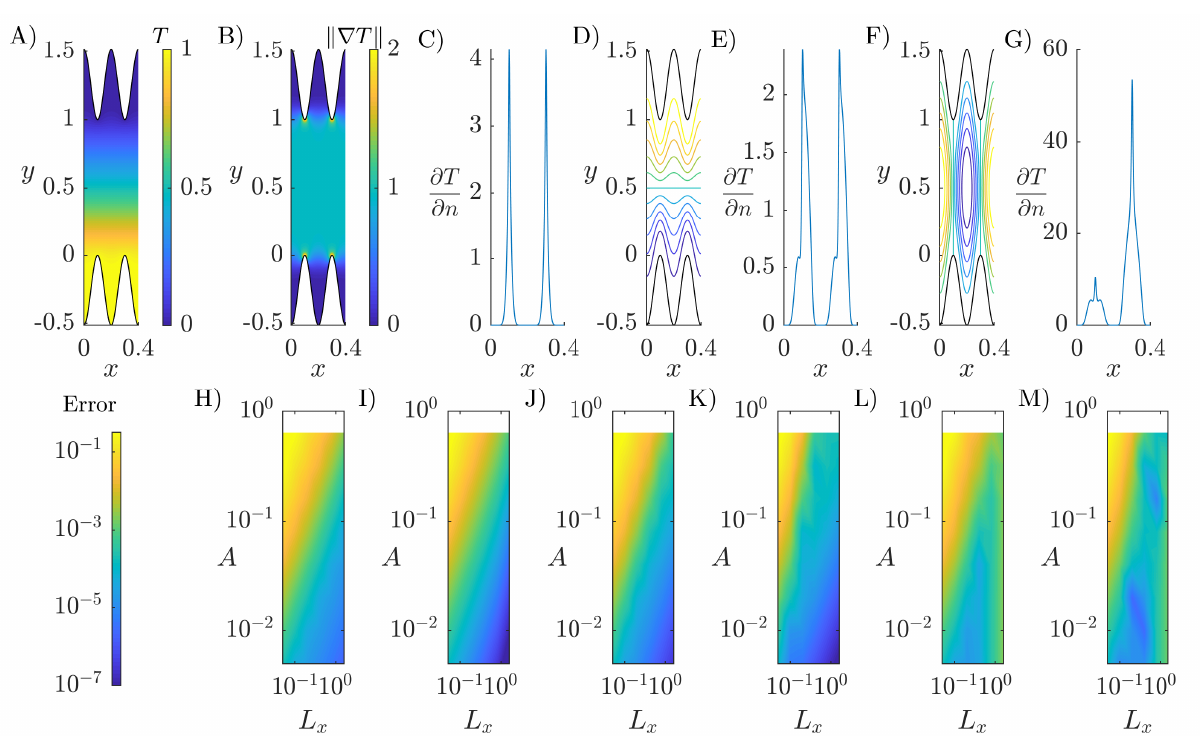}
    \caption{\footnotesize Examples of the temperature fields and heat flux near wavy walls, and corresponding relative errors. For sinusoidal wavy walls with 
    $A$ = 0.4 and no flow, A) the temperature field, B) the norm of its gradient, and C) the heat flux density along the bottom wall. D) Streamlines for a steady flow through the wavy channel with Pe = 10$^4$. E) Heat flux density along the bottom wall corresponding to the flow in panel D. F) Streamlines for steady convection rolls with Pe = 10$^4$. G) Heat flux density along the bottom wall corresponding to the flow in panel F. H-M) Relative errors in $\partial T/\partial n$ and Nu, computed on a 256-by-257 mesh relative to a 512-by-257 mesh, for various choices of wall perturbation amplitude $A$ and horizontal period $L_x$. Specifically, H and I plot the maximum relative errors in $\partial T/\partial n$ along the bottom wall and the relative error in its mean, for the case of no flow (corresponding to panels A-C). We plot the same error quantities for the flow through the channel (panel D) in panels J and K, and for the convection rolls (panel F) in panels L and M. The colorbar at the bottom left shows the error values.}
    \label{fig:DTDnResPoisRollsFig}
             \vspace{-.10in}
\end{figure}

It turns out to be computationally challenging to compute solutions accurately with both large wall-perturbation amplitudes $A$ and with many modes (large $M_1$), or with small $L_x$. This challenge occurs at small $L_x$ even though we use a horizontal coordinate $p$ that is $x$ scaled by $L_x$, so the number of grid points per period is fixed. The challenge is illustrated numerically by considering the simplest case of no flow, i.e. Laplace's equation with wavy walls. The top row of figure \ref{fig:DTDnResPoisRollsFig} shows temperature and heat flux distributions in this and two other simple cases. The bottom row shows measures of the numerical errors in these cases with 256 grid points uniformly spaced in the $p$ coordinate. 

To begin, figure \ref{fig:DTDnResPoisRollsFig}A shows the pure diffusion temperature field for $A$ = $L_x$ = 0.4. The temperature gradient is nearly uniform in the central region, $0 \leq y \leq 1$, and nearly zero above and below this region. Panel B shows this clearly by plotting values of the temperature gradient norm, which transitions from 1 in the central region to 0 in the remainder of the domain. However, there are small regions at the wall inward extrema (near $y = 0$ and 1) where the gradient norm rises sharply to slightly above 4. Panel C plots the wall heat flux along the bottom wall, which has sharp peaks at the wall's inward extrema. The wall shape is sinusoidal with two periods per domain period $L_x$, and is easy to resolve with 256 points per $L_x$. But the resulting heat flux peaks are much sharper and more difficult to resolve. For Laplace's equation there are many alternative methods, such as boundary integral methods, that would give better resolution of the boundary heat flux with all the grid points concentrated along the boundary. However, in general we have the steady advection-diffusion equation with arbitrary flow fields in the advection-dominated limit, which requires a fine grid in the interior, not just on the boundary. Panels D and E show the streamlines and boundary heat fluxes for a second case, a Poiseuille-like flow that conforms to the boundary, $\psi = c(3q^2-2q^3)$, with $c$ chosen so Pe = 10$^4$. Panel E shows that the peaks in $\partial_n T$ decrease and become asymmetrical, but are still quite sharp. Panels F and G show a third case, $\psi = c(3q^2-2q^3)\cos{2\pi p}$, with $c$ again chosen so Pe = 10$^4$. The streamlines (panel F) show that we have chosen convection rolls aligned with the walls' extrema, and that between the convection rolls, vertical jets of fluid run from one wall extremum to the other. The configuration is similar to some of the optimal flows and wall shapes that will be presented later. The heat flux from the bottom wall (panel G) is greatly increased from the previous cases at its maximum, where the downward jet impinges on the lower boundary, as well as at the smaller peak where the upward jet emanates from the lower boundary. All three cases are qualitatively similar in having sharp peaks in $\partial_n T$ that require a fine grid to resolve.

The bottom row plots estimates of the errors in $\partial_n T$ in the three cases. For each case, two measures of error are used: the max norm ($\|\cdot\|_\infty$), and the relative error in Nu. The error is estimated by comparing $\partial_n T$ at the lower wall, computing $T$ with a 512-by-257 ($p,q$) grid and the 256-by-257 grid that omits every other point in the $p$ direction. Panels H and I show
\begin{align}
    \Delta_{rel}\|\partial_n T\|_{\infty} \equiv \frac{\max_p{\left|\partial_n T_{256}(p,0) - \partial_n T_{512}(p,0)\right|}}{\max_p\left|\partial_n T_{512}(p,0)\right|} \; ; \;
    \Delta_{rel} \mbox{Nu} \equiv \left|\frac{\mbox{Nu}_{256} - \mbox{Nu}_{512}}{\mbox{Nu}_{512}}\right| \label{Err}
\end{align}
respectively for the case of no flow, panels A-C. Panels J/K and L/M show the same quantities for the second and third cases (panels D/E and F/G respectively). In each panel in the bottom row, the errors have a similar distribution in $L_x$-$A$ space. Errors are largest at small $L_x$ and large $A$, and are roughly constant along curves $A = L_x^r$ for $r$ slightly larger than 1.
The $\Delta_{rel}\|\partial_n T\|_{\infty}$ errors (panels H, J, and L) are significantly larger than the corresponding $\Delta_{rel} \mbox{Nu}$ errors (panels I, K, and M), particularly in the second and third cases (J versus K and L versus M). This is perhaps not surprising since $\Delta_{rel}\|\partial_n T\|_{\infty}$ is a measure of local error and is more sensitive to how well the peak of $\partial_n T$ is resolved. 

Our optimization algorithm computes Nu over a much wider range of flows than the three cases here, but when we check the accuracy of Nu for our optimal flows, the same general trend holds: for a given grid size and a given level of error, there is a limit to how large $A$ can be and how small $L_x$ can be. We also find large errors if we exceed relatively small integer values of $M_1$, the number of modes that describe the wall shape. Increasing $M_1$ allows for finer length scales in the wall shape, similarly to decreasing $L_x$, except that the grid spacing is proportional to $L_x$ but does not change with $M_1$. In this study we perform the optimization with $M_1$ ranging from 1--4, with a smaller maximum $A$ required at larger $M_1$. The limit on $A$ can be (very roughly) approximated by a power law of the form $M_1^R$ with $R \approx -1.7$. Fortunately, we find in our optimal flow computations that the optimal flows give significantly lower relative errors at a given ($A$, $L_x$) pair than the values shown in figure \ref{fig:DTDnResPoisRollsFig}H--M. The reason is not entirely clear, but the lower errors are seen across a wide range of different optimal flow and wall configurations. Specifically, at Pe = 10$^2$, 10$^3$, \ldots, 10$^7$, we take the 2--4 flow and wall shape optima with largest Nu at each $M_1$ = 1-4, yielding 10--12 optima in total at each Pe. The optima are computed with $m$ = 256 and $n$ ranging from 256 to 1024, with larger $n$ at larger Pe. For each case, we compute the error quantities (\ref{Err}) by doubling $m$ and keeping $n$ fixed. We find that $\Delta_{rel} \mbox{Nu} \leq 0.012$ across all 78 optima, and $\leq 0.005$ in 74 out of 78 cases. $\Delta_{rel}\|\partial_n T\|_{\infty} \leq 0.13$ in all 78 cases with a median value of 0.044. The maximum error occurs at a sharp peak of $\partial_n T$ and is much smaller elsewhere, which is why the error in Nu is much smaller. Since Nu is our main focus of interest, $m$ = 256 is a reasonable value to use for the optimization, since much larger $m$ would greatly slow the computations.

We also study the analogous error quantities when $n$ is doubled and $m$ is fixed. The maximum of $\Delta_{rel} \mbox{Nu}$ is somewhat larger
than previously, 0.033 over the 78 cases, though it is below 0.0076 in 74 out of 78 cases. By contrast, the maximum $\Delta_{rel}\|\partial_n T\|_{\infty}$ is much smaller than previously, 0.041 over the 78 cases, with a median value of 0.0021.

Having described the computational framework, we now use it to validate the decoupled approximation (\ref{NuExpn2}) that was used in section \ref{sec:SmallPe} to explain the transition to optima with nonflat walls at a critical Pe $>$ 0.

\section{Testing the accuracy of the decoupled approximation \label{sec:TestDecoupled}}

\begin{figure}[ht]
    \centering
    \includegraphics[width=5.5in]{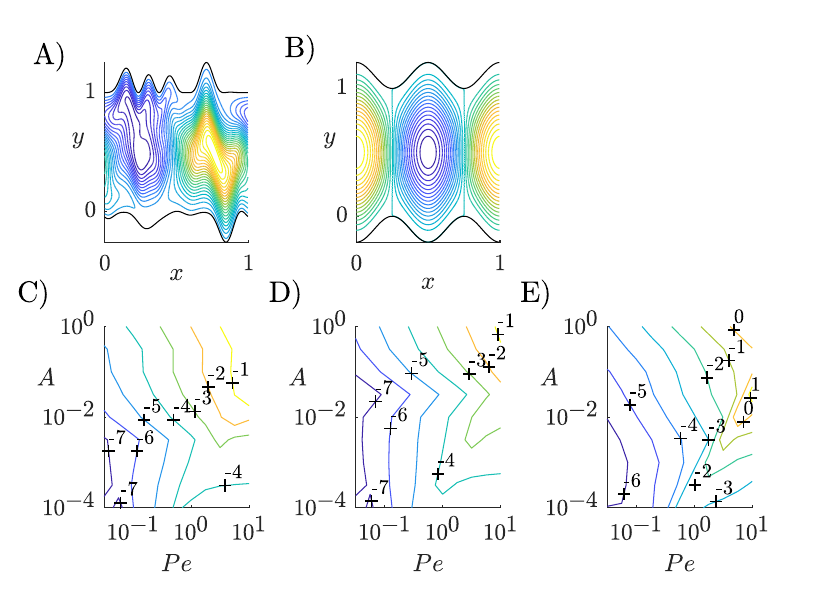}
    \caption{\footnotesize Validation of (\ref{NuExpn2}) which shows the decoupled effect of small wall and flow perturbations on Nusselt number at leading order. A) Example of a wall shape and flow streamlines using randomly-generated coefficients. B) Streamlines of the optimal flow for Pe = 10$^{1.5}$ (with flat walls), stretched vertically to correspond to a sinusoidal wall perturbation. C and D) Contour maps of the relative errors (log base 10) in the linear decoupled approximation to the Nusselt number for the streamlines and wall shapes in panels A and B, but with a range of flow and wall perturbation amplitude Pe and $A$, respectively. The relative error is $\log_{10} (|$Nu-(Nu$_A$+Nu$_B$-Nu$_0)|/|$Nu-Nu$_0|)$, with the quantities as defined in the main text.
    E) Contour maps of the maximum of the relative error over the values in C, D, and an ensemble of 50 other cases described in the main text.}
    \label{fig:SmallPeAndMaxAmpFig}
             \vspace{-.10in}
\end{figure}

In section \ref{sec:SmallPe} we derived an expansion for Nu for small flow and wall perturbations in which the leading-order effects of the perturbations are decoupled.
In figure \ref{fig:SmallPeAndMaxAmpFig} we show that
the expansion (\ref{NuExpn2}) is accurate for a large ensemble of flows and walls shapes over ranges of small Pe and $A$.
The expansion should hold for any flows and wall shapes---not just optima---as long as Pe and $A$ are sufficiently small.

To test this hypothesis, we take $m = n = 256$, set $M_1 = 4$ giving 9 modes for each wall shape, and use 15 flow modes (products of the first three modes in the $p$ direction with the first five modes in the $q$ direction, corresponding to $M = 1$ and $N$ = 8). We generate 50 random sets of coefficients for all the modes to obtain an ensemble of 50 sets of random wall shapes and flows (one example is shown in figure \ref{fig:SmallPeAndMaxAmpFig}A, with $A = 10^{-0.5}$), and add two more nonrandom choices. These are the optimal flows at Pe = 10$^{1.5}$ and 10$^2$ with flat walls, i.e. convection rolls $\psi(p,y)$ on a rectangular domain $0 \leq p, y \leq 1$. We set $y_{bot}(p)$ and $y_{top}(p)$ to sinusoidal walls of varying $A$, and substitute $q$ for $y$ in $\psi$, essentially stretching the rectangular convection rolls to fit the wavy walls. Figure \ref{fig:SmallPeAndMaxAmpFig}B shows an example---the optimal flow at Pe = 10$^{1.5}$ with $A = 10^{-0.5}$. 

For the resulting set of 52 flow and wall-shape combinations, we vary Pe from 10$^{-1.5}$ to 10$^1$, vary $A$ from 
10$^{-4}$ to 10$^0$, and compute Nu in each case.
In order to validate (\ref{NuExpn2}) we need to compute
Nu$_{1A}$ and Nu$_{1B}$ and compare the sum with Nu - 1.
We do this in a way that highlights the decoupled effect of the flow and wall perturbations. Nu with both the flow and wall perturbations should be approximately Nu$_0$ + Nu$_{1A}$ + Nu$_{1B}$. Nu with the same wall perturbation but no flow---call it Nu$_A$---should be approximately Nu$_0$ + Nu$_{1A}$. Nu with the same flow perturbation but no wall perturbation---call it Nu$_B$---should be approximately Nu$_0$ + Nu$_{1B}$. Summing the last two expressions, Nu$_A+$Nu$_B-$Nu$_0$ should be approximately Nu. In figure \ref{fig:SmallPeAndMaxAmpFig}C--E we plot the absolute value of the ratio of the difference of Nu and Nu$_A+$Nu$_B-$Nu$_0$---which should be the omitted higher-order terms in (\ref{NuExpn2})---relative to Nu$-$Nu$_0$, which approximates the first-order terms. If the ratio is small, the first-order terms that arise from decoupling the wall and flow perturbations are indeed dominant over the remaining terms. Panel C gives the relative error (log base 10) for the flow and wall configuration in panel A, panel D gives the relative error for panel B, and panel E gives the maximum of the relative errors in panels C, D, and the 50 other flow and wall shape combinations in the ensemble of 52 described earlier. Panels C--E show that the error in the decoupled approximation is indeed small when $A$ and Pe are small.
Panel E shows the relative errors become significant in some cases when Pe reaches 10$^{0.3}$. The transition to optima with wavy walls occurs at higher Pe, between 10$^{1}$ and 10$^{2}$. 

It is possible that the largest relative errors in panel E are due to flows that are unlike the optimal flows and wall shapes near the transition. We will see in figure \ref{fig:Power1e4Fig} that near the transition the optima are relatively smooth, as in panel B, and compared to the less smooth case in panel A, we can see smaller relative errors in the decoupled approximation for the smoother case (contours of a given value in panel D are shifted upwards and rightwards relative to those in panel C). Therefore, for the optimal flows, perhaps the decoupled approximation remains accurate up to higher Pe than for some of the random flows studied here.







\section{Moderate-to-large convection (Pe $\gtrsim$ 1) \label{sec:LargePe}}

\begin{figure}[h]
    \centering
    \includegraphics[width=5in]{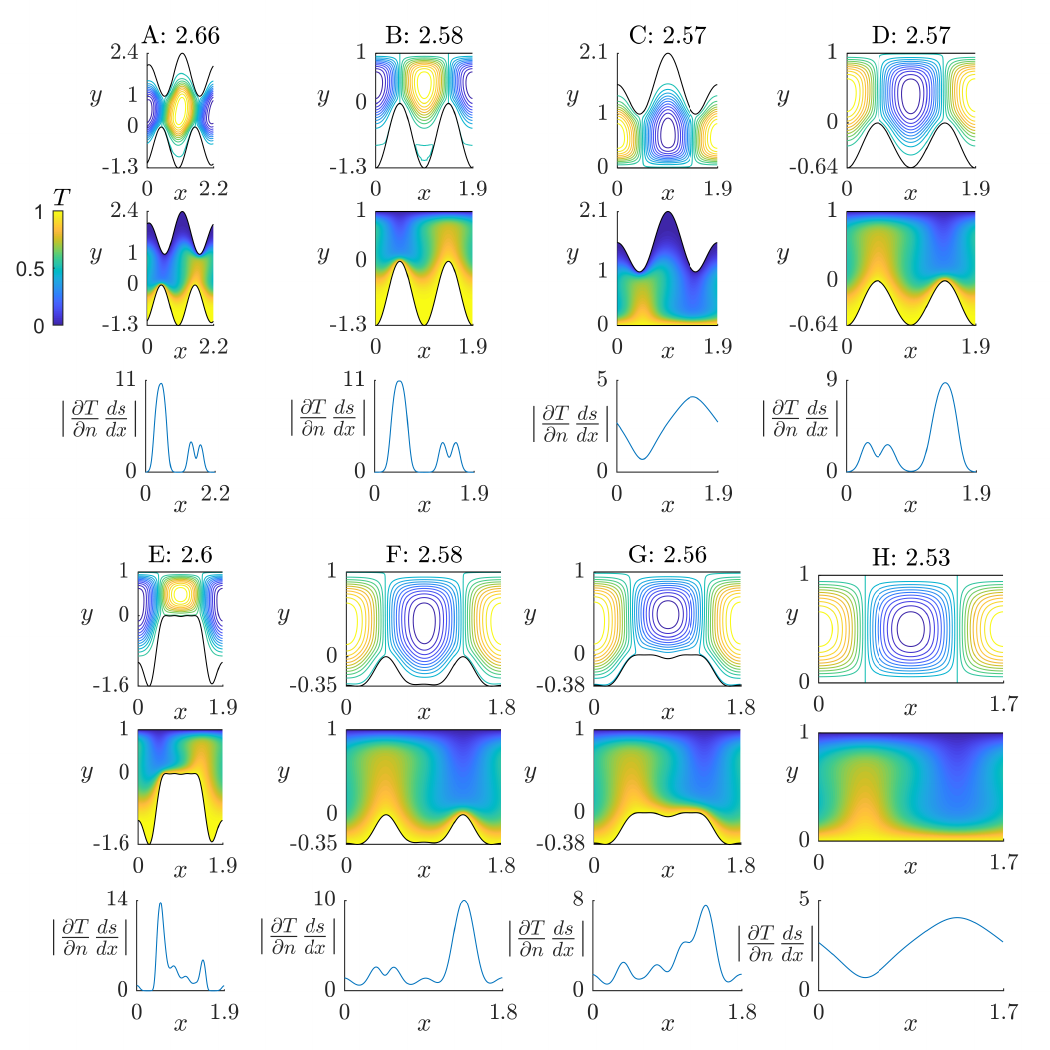}
    \caption{\footnotesize Optimal flows and wall shapes, temperature fields, and heat flux densities at Pe = 10$^2$. Eight cases are shown, four in the top three rows and four in the bottom three rows. Each case is labeled A-H at the top, followed by the Nusselt number value. Below the label are three rows with: a contour plot of the streamlines (top row), the temperature field color plot (middle row, with color bar at left), and a plot of the heat flux distribution versus the horizontal coordinate along the lower wall (bottom row). The numbers of modes describing the walls' shapes ($M_1$) are (A, B) 1, (C, D) 2, (E-G) 3, and H shows the flat-wall optimum for comparison.}
    \label{fig:Power1e4Fig}
             \vspace{-.10in}
\end{figure}

\begin{figure}[h]
    \centering
    \includegraphics[width=5in]{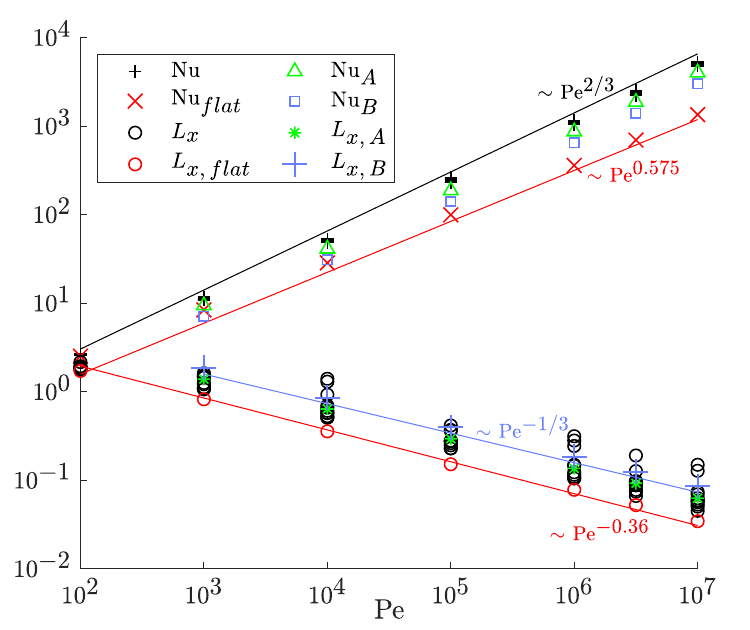}
    \caption{\footnotesize Values of Nu and $L_x$ for optima with nonflat walls (black plusses and circles, respectively), with flat walls (red crosses and circles, respectively), and for two special flow and wall configurations A and B (green and blue symbols, respectively), for Pe from 10$^2$ to 10$^7$.}
    \label{fig:NuLxPlot}
             \vspace{-.10in}
\end{figure}

The transition to optima with nonflat walls at a critical nonzero Pe value, predicted theoretically in section \ref{sec:SmallPe}, is consistent with our optimization calculation. We find only optima with flat walls for Pe $\leq 10^1$, but at Pe $\geq 10^2$, we find optima with nonflat walls that outperform those with flat walls---by an increasing margin as Pe increases, we will show. Using plots of Nu versus Pe (like figure \ref{fig:NuLxPlot}, shown later) for the best local optima with flat and nonflat walls, we find the two curves meet at Pe $\approx 10^{1.7}$, which approximates the transition Pe value where nonflat walls become optimal.

For Pe $= 10^2$, the top-performing optima are shown in figure \ref{fig:Power1e4Fig}A--G, and the flat wall case, also a local optimum, is shown in panel H. For each optimum, three panels are shown. The top panels, labeled A--H and followed by the Nu values, show the streamlines and wall shapes. The middle panels show the temperature fields, and the bottom panels show the heat flux per unit horizontal width along the bottom walls. These optima have
different numbers of modes ($M_1$) for the walls' shapes---(A, B) 1, (C, D) 2, (E-G) 3---and the Nu values, shown at the top, do not correlate strongly with $M_1$. 

The convection rolls with nonflat walls (A--G)  roughly resemble those with flat walls (H), but are stretched by various amounts vertically, either upward, downward, or both. The nonflat cases have slightly larger $L_x$ values, 1.9--2.2, versus 1.7 for the flat case.  

The temperature fields (middle panels) are generally similar in all cases, with one upward hot plume and one downward cold plume. Cases A--C and E have large wall deformations but near the peaks of the deformations there is almost no flow, and the temperature is almost constant. These regions could be extended farther downward (for the bottom wall peaks) or upward (for the top wall peaks) without significantly altering the flow and temperature elsewhere, the net power dissipation, or the net heat transfer. Therefore, there is some insensitivity to wall perturbations beyond a certain magnitude. 

The distributions of heat flux per horizontal width from the bottom wall (bottom panels) clearly show the effect of waviness in the bottom wall (A,B, D--G) compared to the cases with flat bottom walls (C, H). In the flat cases, there is a peak in heat flux where the cold plume reaches the bottom wall, a trough where the hot plume emerges from the wall, and a monotonic change between them. For the nonflat walls, there is one large peak where the downward jet impinges on an upward peak in the bottom wall. There is usually a smaller ``dimpled" peak
(in A, B, D, and F) where the upward jet emanates from another upward peak on the bottom wall.
Between these peaks, the rest of the heat flux distribution is essentially zero, in the troughs along the bottom wall. These are in the aforementioned ``dead zones" of no flow and nearly constant temperature corresponding to nearly zero heat flux density. In each of these cases,
the upward peaks of the bottom wall usually align with the upward and downward jets, i.e. the streamlines that connect stagnation points on the top and bottom walls. Although the heat flux distributions are quite different from panel H, the net increases in Nu are modest, about 1-5\% at this Pe.  Panels E and G show a similar but less symmetrical changes in the extrema in the heat flux distributions, because the upward peaks of the bottom wall are asymmetrical, sloped on one side (where heat transfer density is larger) and flattened on the other (where heat transfer density is smaller).

\begin{figure}[h]
    \centering
    \includegraphics[width=6in]{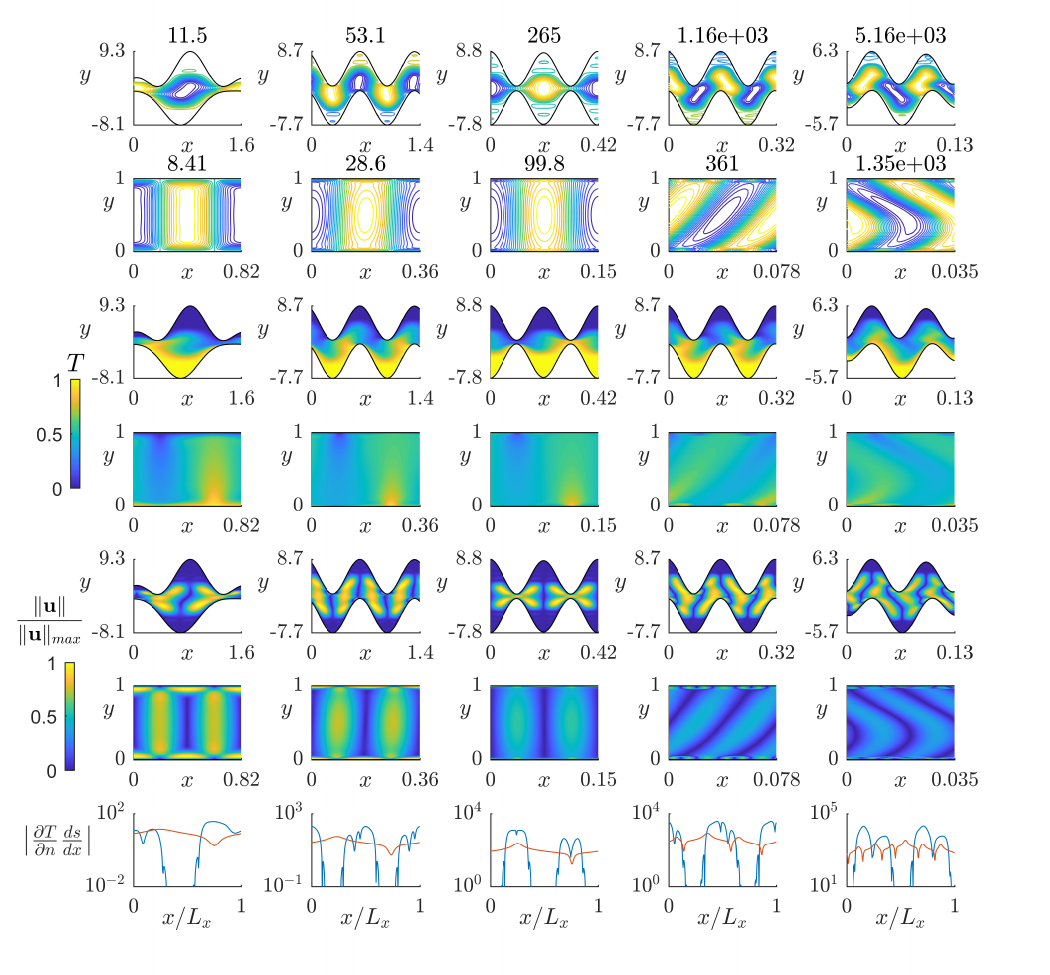}
    \caption{\footnotesize Comparisons of the best optima with nonflat walls and flat walls. Each column compares the two cases for Pe = 10$^3$, 10$^4$, 10$^5$, 10$^6$, and 10$^7$ (in order from left to right). In each column, the top two rows show streamline contour plots for the optima with nonflat and flat walls respectively, each labeled at the top by the corresponding Nu value. The third and fourth rows show the temperature fields (color bar at left). The fifth and sixth rows show the flow speed distributions. The seventh row compares the heat flux distributions along the bottom wall for the optima with nonflat walls (blue) and flat walls (red).}
    \label{fig:BestVsFlatFig}
             \vspace{-.10in}
\end{figure}

Next, we compare the best optima with nonflat and with flat walls across a range of Pe extending up to five orders of magnitude larger than figure \ref{fig:Power1e4Fig}.  Each column of figure \ref{fig:BestVsFlatFig} compares these two optima at Pe = 10$^3$, 10$^4$, 10$^5$, 10$^6$, and 10$^7$ (in order from left to right). In each column, the top six panels arrayed vertically show the streamlines (first and second rows, Nu values at the top), temperature fields (third and fourth rows, color bar at left), flow speed distributions (fifth and sixth rows, color bar at left), with the nonflat optima above the flat optima in each case. The seventh row shows the bottom-wall heat flux distributions (flat-wall case in red, non-flat-wall case in blue). 

First, we note from the Nu values at the top of the streamline plots (top two rows) that Nu is larger for the nonflat walls, 37\% larger at Pe = 10$^3$ and a factor of 3.8 larger at Pe = 10$^7$. The streamline distributions show a series of convection rolls for both flat and nonflat walls. Unlike in figure \ref{fig:Power1e4Fig}, the $x$ axes are dilated relative to the $y$ axes. This makes the flow structures visible at large Pe and in all cases with nonflat walls, whose deflections are very large. The streamlines are somewhat concentrated away from the walls in the nonflat cases, leaving ``dead zones" near the walls. The nonflat cases' streamlines show a few different types of configurations. The first and third columns have convection rolls positioned horizontally between the walls' inward extrema, though the two rolls are much more symmetrical in the third column. In the second column, the convection rolls touch the inward extrema on one side only, and lie below the {\it outward} extrema on the other side. The fourth and fifth columns' rolls are similar but skewed, so they meet the inward extrema at an oblique angle. We will show later that similar versions of these different optima and many more seem to occur in all Pe with wavy walls. That is, the same types of optimal flow and wall configurations occur at all Pe $\gtrsim 10^3$. The main difference is simply that the horizontal period $L_x$ decreases with Pe as a power law (to be shown later). Both walls are wavy at these higher Pe, unlike for most cases at $10^2$ in figure \ref{fig:Power1e4Fig}.

The corresponding flat-wall optima, shown in the second row, have predominantly rectangular convection rolls with $L_x$ that also decreases as a power law. The $L_x$ values in the flat cases are about a factor of 4 smaller than in the nonflat cases at the largest Pe, but the difference is closer to a factor of 2 when one considers that there are two pairs of convection rolls in the top row of the fourth and fifth columns. There is a strong skewness in the flat walls' convection rolls in the last two columns, but this is exaggerated by the $x$-dilation, and does not seem to influence heat transfer much \cite{alben2023transition}.
Actually, there is a key difference in the flat-wall optima that is difficult to see in the streamline plots. They have a thin boundary layer whose thickness scales as Pe to a power, close to that of $L_x$. The boundary layer switches from a horizontal flow along the wall to branching flows in the last two columns, too thin to be seen clearly. No similar boundary layer is seen in the wavy-wall optima.

The presence or absence of boundary layers is somewhat apparent in the temperature fields, shown in the third and fourth rows. In the third row, as in figure \ref{fig:Power1e4Fig}, there are hot and cold temperature plumes aligned with the downward and upward jets that run between the walls. The full range of temperatures from 0 to 1 can be seen clearly in all the columns of the third row, across Pe.  The fourth row also has hot and cold temperature plumes, but moving leftward toward larger Pe, the temperature values are more concentrated in the middle range, near 0.5. Temperatures close to 0 and 1, the wall temperatures, are confined to ever thinner boundary layers. The flow speed fields, in the fifth and sixth rows, show the boundary layer phenomenon as well. In the fifth row, with wavy walls, the full range of flow speeds occurs throughout the domain and is equally visible at small and large Pe. In the sixth row, by contrast, the largest flow speeds (yellow) are less visible at large Pe because they are more confined to the boundary layer. Evidence for boundary layers is not particularly evident in the seventh row, the bottom-wall heat flux distributions, but clear differences are seen between the wavy-wall and flat-wall distributions (blue and red, respectively). Log scales are used instead of the linear scales in figure \ref{fig:Power1e4Fig}, but otherwise the flat-wall distributions are similar, with intervals of monotonic change, more numerous for the branching flows in the fourth and fifth columns. The wavy-wall distributions show peaks near the convection rolls and decay sharply in the ``dead zones" where temperatures are nearly constant.

In figure \ref{fig:NuLxPlot} we plot Nu and $L_x$ (red crosses and circles) for the best flat-wall optima at Pe = 10$^2$, 10$^3$, \ldots, 10$^6$, 10$^{6.5}$, 10$^7$, and at each Pe, Nu and $L_x$ values (black plusses and circles) for an ensemble of 8-12 nonflat optima that include the 2-4 best optima
at each $M_1$ from 1 to 4, including all the cases in figures \ref{fig:Power1e4Fig} and \ref{fig:BestVsFlatFig}. The fit lines showing Nu $\sim$ Pe$^{0.575}$ and $L_x \sim$ Pe$^{-0.36}$ are shown in red next to the corresponding flat-wall data. There are subtle but noticeable changes in the positions of the data relative to the fit lines when Pe exceeds the branching transition value for the flat-wall optima, between Pe = $10^5$ and $10^{5.25}$ \cite{alben2023transition}. At lower Pe, the flat-wall Nu are better fit by Pe$^{0.54}$, while the $L_x$ values (red circles) lie slightly below the red fit line versus slightly above at larger Pe. The Nu values for the nonflat optima (black plusses) seem to follow a single trend, Pe$^{2/3}$, more consistently. This is the scaling that was found for computations of 3D optimal flows with {\it flat} walls by \cite{motoki2018maximal}. This is also an a priori upper bound that was proved using the background method for the 2D flat-wall case \cite{souza2016optimal}, and for the same class of nonflat walls considered here by Goluskin and Doering \cite{goluskin2016bounds}.

\begin{figure}[h]
    \centering
    \includegraphics[width=5in]{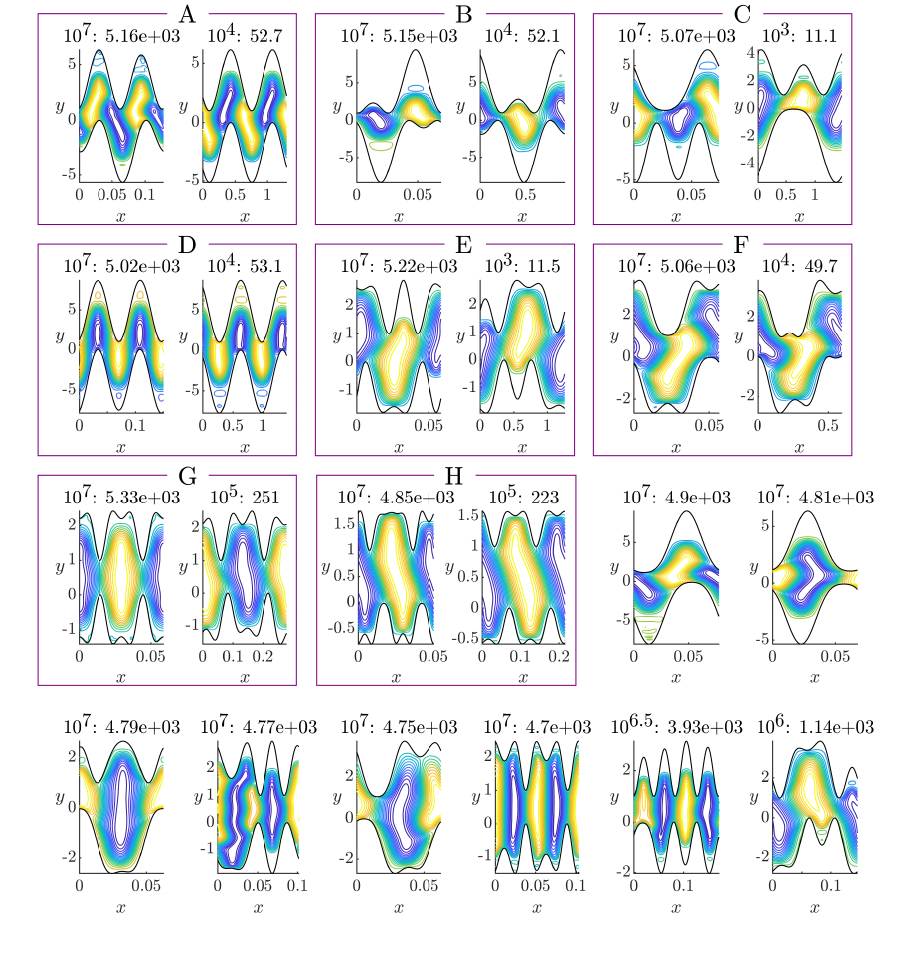}
    \caption{\footnotesize Typical flow and wall configurations of optima. Each purple box (labeled A-H at the top) shows a pair of streamline contour plots for optima with similar flow and wall configurations, one at high Pe (left) and one at low Pe (right). At the top of each contour plot is a label showing (Pe value): (Nu value). Following the eight pairs are eight individual examples of other optimal flows and wall shapes (the two rightmost plots in the third row and all six of the plots in the fourth row).}
    \label{fig:OptimaLowHighPeFig}
             \vspace{-.10in}
\end{figure}

There is much more scatter in the corresponding $L_x$ values (black circles), partly because for these optima the horizontal periods may contain one or two pairs of convection rolls. Differences in flow and wall configurations like in the top row of figure \ref{fig:BestVsFlatFig} also account for a significant amount of the scatter in $L_x$ values, as can be seen subsequently in figure \ref{fig:OptimaLowHighPeFig}, by comparing $L_x$ for flows at the same Pe. The $L_x$ values in the flat cases scale approximately as Pe$^{-0.36}$, while in the nonflat cases, the bottom envelope of the black circles scales approximately as Pe$^{-0.35}$. These scalings are much closer than those for the Nu values.

By extrapolating the plots of Nu versus Pe for flat and nonflat walls to smaller Pe, we find the two curves meet at Pe $\approx 10^{1.7}$, which approximates the transition Pe value where nonflat walls become optimal.

In figure \ref{fig:NuLxPlot} we also plot Nu and $L_x$ values (green and blue, labeled A and B in the legend) for two specific flow configurations to be discussed later.

When discussing the best optima with wavy walls that we found, shown in figure
\ref{fig:BestVsFlatFig}, we suggested that all the optima are different members of a single class that occurs at all Pe $\gtrsim 10^3$, and are simply scaled by different $L_x$ at each Pe. In figure \ref{fig:OptimaLowHighPeFig} we provide evidence for this claim. Starting from the top, we have eight purple boxes, labeled A--H, showing a pair of optima. The first occurs at a large Pe ($10^7$, labeled at the top), and the second at a much smaller Pe, $10^3$--$10^5$. The two are chosen from among the four top optima at each $M_1$ ranging from 1 to 4 (except for the second member of pair H which is eighth best at Pe = 10$^5$ and $M_1 = 3$). In A--D, $M_1 = 1$; in E and F, $M_1 = 2$; in G and H, $M_1 = 3$ except for the first member of pair H, which has $M_1 = 4$. The remaining flows, following the eight pairs, are chosen simply for their differences from the preceding flows and from each other, to show the diversity of flows and wall configurations that occurs among the optima. 

Pair A are optima that show essentially the same pattern of streamlines and wall shapes (but with reversed flow directions), even though Pe differs by a factor of $10^3$, Nu by a factor of about $10^2$, and $L_x$ by about a factor of 10. Pair B has streamlines that are more confined vertically than pair A, and with more asymmetric walls, but very similar Nu values, showing that very different flows can achieve heat transfer close to the top values we have found. Pair C have a more complicated wall shape and less symmetric vortices than pairs A and B. The second member of pair C is approximately obtained by reflecting the first member about a horizontal line and reversing the flow direction, and increasing $L_x$ by a factor of about 20. Pair D is an almost identical pair of wall shapes and flows, except for the difference in $L_x$. 
In E and F, $M_1$ is increased to 2, allowing for somewhat more complicated wall shapes and vortices. Pairs G and H have wall shapes that commonly occur at $M_1 = 3$ and 4, with peaks resembling indented hats. Pairs A--H are just a small sample of the similar flows and wall shapes that occur across $10^3 \leq$ Pe $\leq 10^7$. Examination of a larger set of 100 optima at each Pe 
= 10$^2$, 10$^3$, \ldots, 10$^6$, 10$^{6.5}$, and 10$^7$ shows that most flow and wall configurations recur throughout this Pe range with little change apart from $L_x$.

The last eight flows and wall shapes in figure \ref{fig:OptimaLowHighPeFig} are chosen to illustrate optimal configurations that are clearly different than those that have already been shown. In order from left to right and top to bottom, these have $M_1$ = 1, 1, 2, 2, 2, 3, 2, and 2, respectively. The convection rolls appear somewhat oblique and asymmetrical in some cases, but when the horizontal dilation of the figures is taken into account---i.e. that $L_x \ll 1$ in most cases---the convection rolls are seen to be very thin horizontally and elongated vertically. The walls have various positions but in all cases the sides of the convection rolls run alongside the walls vertically for O(1) distances at all Pe, allowing for heat transfer over a large surface length. 

\begin{figure}[h]
    \centering
    \includegraphics[width=5.5in]{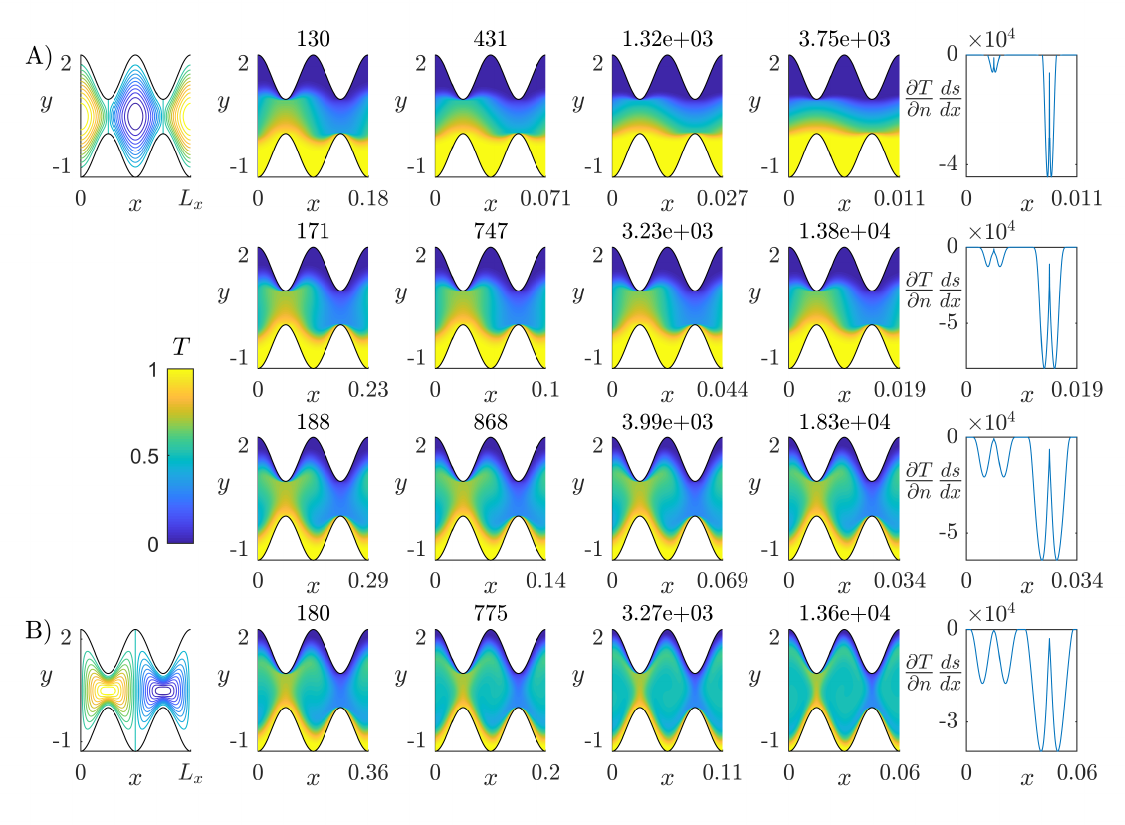}
    \caption{\footnotesize For the flow and wall configuration in panel A, variations of temperature fields and heat flux distributions with $L_x$ are shown at four large Pe values. The same streamline pattern, shown in the upper left panel, is used in all cases. In the four-by-four grid of temperature fields, Pe increases from left to right with the values 10$^5$, 10$^6$, 10$^7$, and 10$^8$. Within each column, Pe is fixed and $L_x$ increases from top to bottom, with the values given at the right ends of the $x$ axes. The column at the far right gives the heat flux distribution along the bottom wall for the rightmost temperature distributions, Pe = 10$^8$. Panel B shows an alternative flow configuration, described in the main text.}
    \label{fig:ScalingsNuLxRollsFig}
             \vspace{-.10in}
\end{figure}

Having presented a variety of optimal flow and wall configurations, we now attempt to understand the basic scalings of the optima with Pe, i.e. Nu $\sim$ Pe$^{2/3}$ (the black plusses in figure \ref{fig:NuLxPlot}) and $L_x \sim$ Pe$^{-0.35}$ (the bottom envelope of the black circles in figure \ref{fig:NuLxPlot}). We consider a simple model flow and wall configuration, the same one we considered in figure \ref{fig:SmallPeAndMaxAmpFig}B, but with $A$ increased to 1. This configuration, shown in figure \ref{fig:ScalingsNuLxRollsFig}A, is a model for some of the optima we have seen, e.g. at the top of the first and third columns in figure \ref{fig:BestVsFlatFig}, and pairs B, G, and H in figure \ref{fig:OptimaLowHighPeFig}. The ``dead zones" of little flow and nearly constant temperature seen in some of those cases can essentially be treated as though the wall were at the inward boundary of the dead zone. There is little viscous dissipation in the dead zone and the temperature at its inward boundary is almost the same as at the wall. To the right of panel A in figure \ref{fig:ScalingsNuLxRollsFig}, a four-by-four array of temperature fields is shown. These result from the configuration in panel A but with the streamfunction scaled to have four different Pe. At each Pe, the flow and wall shape are scaled horizontally to have four different $L_x$. Thus, from left to right, the four columns correspond to Pe = 10$^5$, 10$^6$, 10$^7$, and 10$^8$ respectively. Within each column, four $L_x$ values are used, increasing from top to bottom (listed at the right end of the $x$ axis in each case). The Nu values are shown at the top of each temperature field, and the maximum values occur in the third row. 

Each column shows a transition between two types of temperature fields as $L_x$ increases. First, we note that all the temperature fields have a basic structure similar to those already described in figures \ref{fig:Power1e4Fig} and \ref{fig:BestVsFlatFig}. In each case there is an upward plume of hot fluid at the left and a downward plume of cold fluid at the right, running between the inward extrema of the two walls. In the top row, the temperature fields resemble those with diffusion only (e.g. figure \ref{fig:DTDnResPoisRollsFig}A) in two ways. First, the temperature is almost constant in the regions near the walls' outward extrema. Second, at each $x$ location the temperature field decreases almost monotonically from the bottom wall to the top wall. The temperature gradient is   
close to linear in the middle part of the flow, $0 \leq y \leq 1$, particularly at the largest Pe (the upper right corner, Nu = 3750).
For the rightmost temperature distributions (Pe = 10$^8$), the column at the far right shows the distributions of heat flux at the bottom wall. Along most of the wall, the temperature gradient is nearly zero, but is large close to the walls' inward extrema, so Nu is much greater than the value with diffusion only, 1. However, with further decreases in $L_x$, the temperature field tends to the diffusion-only case and Nu $\to 1$.

In the bottom row at the largest $L_x$, the temperature is more well-mixed in the middle part of the flow, with temperatures close to 1/2. This well-mixed fluid penetrates close to the walls even at their outward extrema, so there is significant heat flux along most of the bottom wall, as shown by the plot adjacent to the bottom-right temperature field. Whereas the top row resembles the diffusion-dominated limit, the bottom row is a more convection-dominated case. As the flow speed increases further, the temperature field in most of the domain becomes closer and closer to 1/2, because the flow on streamlines adjacent to the walls exchanges only a small amount of heat with the hot and cold walls during the short time it passes along either one. The hot upward and cold downward plumes shrink to very thin jets concentrated along the streamlines that emanate from the stagnation points on the upper and lower walls. This extreme is suboptimal for efficiency in two ways. First, viscous power is expended for the flow in the middle of the domain, but it transfers little heat because the temperature is almost uniform there, so the temperature gradient is small. Second, the upward and downward plumes that do transport heat are more widely separated in the bottom row than in the rows above. Therefore there is less heat exchange per unit horizontal width. By comparison, the top row has more upward and downward plumes per unit horizontal width, but the small values of $L_x$ there mean that the flow needs to be much slower in the top row than in the bottom row to have the same viscous dissipation rate. Hence a more viscous-dominated temperature field occurs. Optimal heat transfer, in the third row, seems to occur at the interface of the viscous-dominated and convection-dominated regimes. $L_x$ is large enough, and thus the flow speed is large enough, to allow cold fluid emanating from the top wall to remain relatively cold as it approaches and passes close to the bottom wall, giving a large temperature gradient there (likewise for the hot fluid emanating from the bottom wall, when it nears the top wall). But $L_x$ is also small enough to have many plumes per unit width.  

We consider again the heat flux distribution plots in the rightmost column. The largest peak values occur in the second row, about twice those of the smallest peak values, in the fourth row. However both plots have about the same average, which is Nu (1.38e+04 versus 1.36e+04, shown above the temperature fields to the left). The larger peaks in the second row have much smaller widths than in the fourth row. The third row has almost the peak values of the second row combined with almost the widths of the fourth row, so it is better than both. The ratio in peak values between the third and fourth row is roughly the ratio of $ds/dx$ values (the inverse of the ratio of $L_x$ values), so the greater heat transfer in the third row is due to having larger $ds/dx$ with about the same $dT/dn$, in other words, more surface area of the bottom wall per unit horizontal width.

For each Pe = 10$^3$, 10$^4$,\ldots, 10$^6$, 10$^{6.5}$, and 10$^7$ we find the $L_x$ that maximizes Nu for the flow in figure \ref{fig:ScalingsNuLxRollsFig}A. The $L_x$ values are plotted as green asterisks in figure \ref{fig:NuLxPlot}, and the Nu values as green triangles. We do the same for the flow in figure \ref{fig:ScalingsNuLxRollsFig}B, which has the convection rolls shifted by a quarter period horizontally. This configuration, with the jets running between the farthest points on the two walls, is more different than flow A from the optimal flows we have seen so far. The Nu values, shown as blue squares in figure \ref{fig:NuLxPlot}, are well below those of flow A, but both have the same scaling, $\sim$ Pe$^{2/3}$ (the slopes are within 0.5\% of 2/3 for each of the line segments connecting the data points at the four largest Pe). The $L_x$ data fit a Pe$^{-1/3}$ scaling nearly as well for both flow configurations. Although flows A and B and the more general optima all have Nu $\sim$ Pe$^{2/3}$, there is a significant difference in the prefactors. The highest Nu for the more general optima is about 30\% higher than that of flow A, which is in turn about 30\% higher than that of flow B. Such differences are perhaps not surprising given the differences in the streamlines of the different flows. Furthermore, all these flows approximate $L_x \sim$ Pe$^{-1/3}$. 

To explain these scalings, we note that flows A and B have the form
\begin{align}
 \psi = c\tilde{\psi}(x/L_x, y) = c\tilde{\psi}(p, y)   \label{PsiScaling}
\end{align}
That is, they each have a single pattern of streamlines given by $\tilde{\psi}$ that is scaled horizontally by $L_x$, and with $c$ chosen to give the power dissipation rate Pe$^2$. The optima from the general optimization problem also seem to have this form, as seen in pairs A--H of figure \ref{fig:OptimaLowHighPeFig} for example. 
For a flow of form (\ref{PsiScaling}) the power dissipation constraint (\ref{Power}) is
\begin{align}
   c^2 \int_0^1 \int_{y_{bot}(p)}^{y_{top}(p)} \left(\frac{1}{L_x^2} \partial_{pp} \tilde{\psi} + \partial_{yy} \tilde{\psi} \right)^2 dp\, dy = \mbox{Pe}^2 \label{PowerScaling}
\end{align}
For flows A and B, $y_{bot}$, $y_{top}$, and $\tilde{\psi}$ are fixed with respect to Pe and $L_x$, and the same holds approximately for pairs A--H of figure \ref{fig:OptimaLowHighPeFig}. The size of $L_x$ determines whether one term in parentheses in (\ref{PowerScaling}) is dominant, and we use this to determine the scaling of $c$ in terms of Pe and $L_x$:
\begin{align}
 L_x \ll 1 \rightarrow c \sim \mbox{Pe}\, L_x^2 \;\; ; \;\;  L_x \gtrsim 1 \rightarrow c \sim \mbox{Pe}. \label{cEst}
\end{align}
We can estimate Nu, the rate of heat transfer per horizontal width, in the case of a thin thermal boundary layer on the walls, e.g. in the last row on temperature fields in figure \ref{fig:ScalingsNuLxRollsFig}. In those cases, the walls are nearly vertical except at their extrema. Near the wall, the flow velocity is approximately vertical, and the vertical component is
\begin{align}
    v = -\partial_x \psi = -\frac{c}{L_x} \partial_{p} \tilde{\psi} &= 
    -\frac{c}{L_x} \left(\partial_{p} \tilde{\psi}|_{p = p_{wall}} + (\partial_{pp} \tilde{\psi}|_{p = p_{wall}})p  +O(p^2)\right) \label{pTaylor} \\
    &= -\frac{c}{L_x} \left( (\partial_{pp} \tilde{\psi}|_{p = p_{wall}})\frac{x}{L_x}  +O\left(\frac{x}{L_x}\right)^2\right) \label{xTaylor} 
\end{align}
\nn In (\ref{pTaylor}) we have Taylor expanded $\partial_{p} \tilde{\psi}$ about a point on the wall. By the no-slip condition, $\partial_{p} \tilde{\psi}|_{p = p_{wall}} = 0$, so removing this term and writing $p = x/L_x$ gives
(\ref{xTaylor}). In the advection-diffusion equation (\ref{AdvDiff}) we can follow the classical boundary layer analysis \cite{leveque1928laws,bergman2011fundamentals,shah2014laminar} and
neglect $\partial_{yy}T$ relative to $\partial_{xx}T$ in the thin thermal boundary layer near a vertical wall, because $\partial_{xx}T \sim 1/\delta^2$, where $\delta$ is the thickness of the boundary layer over which the temperature changes by $O(1)$ (from 0 or 1 at the wall to about 1/2 in this problem), while $\partial_{yy}T \sim O(1)$, since the temperature changes by $O(1)$ over an $O(1)$ distance along the wall. Since $u \ll v$ near the wall, $u\partial_x T$ and $v\partial_y T$ are of the same order \cite{leveque1928laws,bergman2011fundamentals,shah2014laminar}.
Using the Taylor expansion (\ref{xTaylor}) and $\partial_{y}T \sim O(1)$,
\begin{align}
   \frac{c}{L_x^2}x \sim v\partial_y T \sim \partial_{xx} T \sim  \frac{1}{\delta^2} \label{BLest}
\end{align}
In (\ref{BLest}) we have $x \sim \delta$ because we are in the thermal boundary layer, so $\delta \sim c^{-1/3} L_x^{2/3}$. The two estimates for $c$ in (\ref{cEst}) then give two estimates for $\delta$,
\begin{align}
 L_x \ll 1 \rightarrow \delta \sim \mbox{Pe}^{-1/3} \;\; ; \;\;  L_x \gtrsim 1 \rightarrow \delta \sim \mbox{Pe}^{-1/3}\, L_x^{2/3}. \label{deltaEst}
\end{align}
One could perhaps obtain more detailed formulae included prefactors using the similarity solution of L{\'e}v{\^e}que \cite{leveque1928laws}, but we focus only on the scaling with Pe here. Nu is given by (\ref{Nu}), and we estimate it assuming the total length of approximately vertical walls per horizontal length $L_x$ is $O(1)$, so
\begin{align}
    \mbox{Nu} &\sim \frac{1}{L_x} \partial_x T \sim \frac{1}{L_x} \frac{1}{\delta} \\
     L_x &\ll 1 \rightarrow \mbox{Nu} \sim \mbox{Pe}^{1/3}\,L_x^{-1} \;\; ; \;\;  L_x \gtrsim 1 \rightarrow \mbox{Nu} \sim \mbox{Pe}^{1/3}\, L_x^{-5/3}. \label{NuEst}
\end{align}
In both estimates in (\ref{NuEst}), smaller $L_x$ gives larger Nu. Therefore we do not consider the possibility that $L_x$ is sufficiently large that the walls are no longer approximately vertical because this case is clearly suboptimal. The assumption of a thin thermal boundary layer constrains how small $L_x$ can be. If $L_x \ll \delta$, the thermal boundary layer is much larger than the horizontal period, i.e. the spacing between approximately vertical segments on the same wall. This is the case in the top row of figure
\ref{fig:ScalingsNuLxRollsFig}, with the smallest $L_x$. In the bottom row, the temperature varies from 0 or 1 to 1/2 over the thermal boundary layer, but in the top row, the boundary layer is so large relative to the spacing between adjacent vertical wall segments that the temperature does not change much from 0 or 1 in the region between the two wall segments. Then there is little heat transfer from the vertical wall segments; it is confined to small regions near the inward wall extrema, also seen in figure \ref{fig:DTDnResPoisRollsFig}C. This is the diffusion-dominated limit, in which Nu is much smaller, O(1) at large Pe. The upper boundary of this regime, $L_x \sim \delta$, is the smallest $L_x$ can be. Thus by (\ref{deltaEst}) and (\ref{NuEst}) we have 
\begin{align}
L_x \sim \delta \sim \mbox{Pe}^{-1/3} \;\; ; \;\; \mbox{Nu} \sim \mbox{Pe}^{2/3}. \label{LxNuPeScalings}
\end{align}

Before concluding, we discuss the observation that the hot and cold wall separation is always close to the minimum, 1, in the computed optima.

\section{Optimality of minimal wall separation \label{sec:MinH}}

We mentioned in section \ref{sec:Model} that the separation distance between the hot and cold walls is constrained to be $\geq 1$, but it turns out to be 1 (to within about 10$^{-4}$) for all the computed optima. We now use a scaling argument to explain why this is so. Let $\{\psi_1(x,y); L_{x,1}; y_{bot,1}(x); y_{top, 1}(x)\}$ maximize
\begin{align}
    \mbox{Nu} = \frac{1}{L_x}\int_{C = \{(x, y_{bot}(x))\}} \partial_n T ds \label{Nu1}
\end{align}
with power
\begin{equation}
    \mbox{Pe}^2 \equiv \frac{1}{L_x}\int_0^{L_x} \int_{y_{bot}(x)}^{y_{top}(x)}  \left(\nabla^2 \psi\right)^2 dy \, dx \label{Power1}
\end{equation}
such that
\begin{align}
    -\nabla^\perp \psi \cdot \nabla T - \nabla^2 T = 0 \label{AdvDiff1}
\end{align}
with temperatures 1 and 0 on the bottom and top walls respectively, and with $\max y_{bot,1}(x) = 0$ and $\min y_{top,1}(x) = 1$. I.e. this is an optimum over all periodic wall shapes with separation distance {\it exactly} 1. Write the resulting values of Nu and the horizontal period as Nu$_1$(Pe),  $L_{x,1}$(Pe). 

Now change the separation distance from 1 to $Y$, any positive value. Let $\{\psi_Y(x,y); L_{x,Y}; y_{bot,Y}(x); y_{top, Y}(x)\}$ maximize Nu with the same power as before, Pe$^2$, and with $\max y_{bot,Y}(x) = 0$, but $\min y_{top,Y}(x) = Y$, so this is the optimum over walls whose separation distance is exactly $Y$ now. If we divide all distances by $Y$, we see that the resulting Nu can be written 
\begin{align}
    \mbox{Nu}_Y = \frac{1}{Y}\frac{Y}{L_{x,Y}}\int_{C = \{x/Y, y_{bot}(x/Y)/Y\}} \partial_n T ds, \label{NuY}
\end{align}
and the power can be written
\begin{equation}
    \mbox{Pe}^2 \equiv \frac{1}{Y^3}\frac{Y}{L_{x,Y}}\int_0^{L_{x,Y}/Y} \int_{y_{bot}(x/Y)/Y}^{y_{top}(x/Y)/Y}  \left(\nabla_{(x/Y),(y/Y)}^2 \psi \right)^2 d(y/Y) \, d(x/Y), \label{PowerY}
\end{equation}
and (\ref{AdvDiff1}) still holds in the rescaled coordinates.
Since $Y$ is fixed, an optimum of problem (\ref{NuY})--(\ref{PowerY}) is also an optimum of problem (\ref{Nu1})--(\ref{Power1}) but with power Pe$^2Y^3$ instead of Pe$^2$. Comparing (\ref{NuY})--(\ref{PowerY}) with (\ref{Nu1})--(\ref{Power1}) we have
\begin{align}
Y \mbox{Nu}_Y = \mbox{Nu}_1(\mbox{Pe} \, Y^{3/2}) \\
L_{x,Y}/Y = L_{x,1}(\mbox{Pe} \, Y^{3/2}).
\end{align}
If we assume power law scalings
\begin{align}
\mbox{Nu}_1(\mbox{Pe}) = a \mbox{Pe}^\alpha \; ; \; 
L_{x,1}(\mbox{Pe}) = b \mbox{Pe}^\beta
\end{align}
then
\begin{align}
\mbox{Nu}_Y(\mbox{Pe}) = Y^{-1+3\alpha/2}\mbox{Nu}_1(\mbox{Pe})  \; ; \; 
L_{x,Y}(\mbox{Pe}) = Y^{1+3\beta/2} L_{x,1}(\mbox{Pe}). \label{NuYLxY}
\end{align}
For the problem of optimizing the flow only but keeping the walls horizontal in \cite{alben2023transition}, we have $\alpha \lesssim 0.575$, at least up to Pe = 10$^7$. Thus $-1+3\alpha/2 < 0$, so $\mbox{Nu}_Y$ decreases as $Y$ increases. In this paper we optimize the flow and the wall shapes, and find asymptotically $\alpha = 2/3$. Thus $-1+3\alpha/2 = 0$, so $\mbox{Nu}_Y$ is independent of $Y$.
The power law behavior is only approximate, however. For flows A and B in figure \ref{fig:ScalingsNuLxRollsFig}, with Nu and the corresponding optimal $L_x$ shown in figure \ref{fig:NuLxPlot}, Nu increases slightly more slowly than Pe$^{2/3}$ but approaches this behavior at large Pe. Therefore, $\mbox{Nu}_Y$ decreases slightly as $Y$ increases, and this is confirmed by comparing Nu$_Y$ for flows A and B with $Y$ = 0.75, 1, and 1.5. For the peak Nu among the computed wavy-wall optima, i.e. the highest black plusses at each Pe in figure \ref{fig:NuLxPlot}, the slopes of the polygonal curve connecting adjacent points fluctuate more strongly about 2/3, from 0.64 to 0.7, than for the Nu data for flows A and B. This may be because the highest black plusses correspond to different flows with different wall shapes and streamline distributions as Pe varies, and thus the Nu values are more strongly affected by the finiteness and randomness of the ensemble of optima. For example, different slopes (but in a similar range, [0.64, 0.69]) are obtained if instead the second-highest black plusses at each Pe are used. However, if the same streamline and wall-shape configuration were used at each Pe, as with flows A and B, we might again obtain a slight decrease in $\mbox{Nu}_Y$ with $Y$, consistent with the wall separation distance always assuming the smallest allowed value, 1. 

As a check, we also compute $L_{x,Y}$ for flows A and B with $Y$ = 0.75, 1, and 1.5, and find very good agreement with $L_{x,Y} \sim Y^{0.5}$, consistent with (\ref{NuYLxY}), with $\beta = -1/3$ from (\ref{LxNuPeScalings}). 

\section{Discussion and conclusions \label{sec:Conclusions}}

We have studied the problem of finding incompressible flows and wall shapes that maximize heat transfer between two walls, one hot and one cold, with a fixed rate of viscous power dissipation for the flow. In order to avoid the singularity when the walls touch, we defined a minimum allowed separation distance between the walls and used it as the characteristic length scale. We showed that with no flow, flat walls maximize Nu. With small flow strength (Pe), we showed that the effect of the walls and the flow approximately decouple, so flat walls are also optimal over some interval of Pe values starting from zero. We then described our computational optimization method, and explained how limits in resolving the temperature gradient relate to limits in the walls' horizontal periods, maximum amplitudes, and the wave numbers of the components defining the walls' shapes. We also verified the accuracy of the decoupled approximation.

At moderate-to-large Pe, we described the transition to optima with wavy walls, and then compared the best optima with wavy and flat walls. The best wavy-wall optima are essentially the same as Pe increases, except for a Pe$^{-1/3}$ scaling of the horizontal period $L_x$. By contrast, the flat-wall optima have an additional boundary-layer structure near the walls, that transitions to branching flows above Pe = 10$^5$. For the optima with wavy walls, Nu scales as Pe$^{2/3}$. We explained the $L_x$ and Pe scalings by studying model flows and showing that the optimal $L_x$ correspond to flows at the interface between the diffusion-dominated and advection-dominated regimes. 

We now briefly explain that the Pe$^{2/3}$ scaling corresponds to an upper bound for all incompressible flows with wavy walls, shown by Goluskin using the background method \cite{goluskin2016bounds}. First, \cite{souza2016optimal} showed that integrating the energy balance equation for Rayleigh-B\'{e}nard convection at a given Rayleigh number Ra over time and space shows that 
\begin{align}
\mbox{Ra}(\mbox{Nu} -1) = \mbox{Pe}^2.  \label{RaNu}  
\end{align}
Goluskin \cite{goluskin2016bounds} used the background method to show that
\begin{align}
\mbox{Nu} - 1 \leq C \mbox{Ra}^{1/2} \label{Ra12}
\end{align}
for Rayleigh-B\'{e}nard convection with rough walls (with $y$ a single-valued function of the horizontal coordinate, as assumed here).
Combining (\ref{RaNu}) and (\ref{Ra12}),
\begin{align}
\mbox{Nu} -1 \leq C\mbox{Pe}(\mbox{Nu} -1)^{-1/2},  
\end{align}
so Nu$-$1 $\leq C$Pe$^{2/3}$, and thus the optimal flow and wall configurations in this work achieve the maximum scaling with Pe for rough walls.

We also used a scaling argument to show that the optima should have the minimum allowed separation between the hot and cold walls, at least for the model flows. These have fixed streamline configurations that are scaled horizontally by the optimal $L_x$ at each Pe, like the computed optimal flows and wall shapes.

We mention briefly now that we also considered an alternative optimization problem, different from what has been studied so far in this paper. We computed solutions that optimize the average rate of heat transfer per unit arc length of the the boundary instead of per unit horizontal length. That is, we replaced $1/L_x$ in (\ref{Nu}) with $1/s_{total}(L_x)$, where
\begin{align}
    s_{total}(L_x) = \int_0^{L_x} \sqrt{1+\frac{dy_{bot}}{dx}^2} + \sqrt{1+\frac{dy_{top}}{dx}^2} dx.
\end{align}
We found that all the optima had nearly flat walls. It is not surprising that with this performance measure, the flat-wall optima would outperform the wavy-wall optima with Nu plotted in figure \ref{fig:NuLxPlot}, at least in the limit of large Pe. The ratio $L_x/s_{total}(L_x) \sim$ Pe$^{-1/3}$, so the Pe$^{2/3}$ scaling in figure \ref{fig:NuLxPlot} would be reduced to Pe$^{1/3}$, while the flat-wall cases would remain the same, $\approx$ Pe$^{0.575}$.

\appendix

\section{Adjoint-based gradient \label{sec:Adjoint}}

Here we give the adjoint-based formulae for the gradient of Nu with respect to the design parameters, $\{\mathbf{c}, \mathbf{B_1},\mathbf{B_2}, A_1, A_2, L_0\}$. For the functions $\{T, \psi, y_{bot}, y_{top}\}$ we define discrete versions $\{\mathbf{T}, \mathbf{\Psi}, \mathbf{y}_{bot}, \mathbf{y}_{top}\}$ which are vectors of values on the appropriate grids in $(p,q)$ or $p$.

Using the discretized version of (\ref{Nu}), Nu is a product of discrete differential and integral operators, i.e. matrices, and $\mathbf{T}$. The matrices depend on $\mathbf{y}_{bot}$, $\mathbf{y}_{top}$, and $L_x$. $\mathbf{T}$ in turn depends on $\mathbf{\Psi}$, $\mathbf{y}_{bot}$, $\mathbf{y}_{top}$, and $L_x$ through the discretized advection-diffusion equation (\ref{AdvDiff}) (including dependences through the discrete differential operators). And our chosen form of $\mathbf{\Psi}$
(\ref{Psi}), which is normalized to satisfy the discretized power dissipation constraint (\ref{Power}), depends on $\mathbf{y}_{bot}$, $\mathbf{y}_{top}$, and $L_x$ through the denominator of (\ref{Psi}).

Thus Nu depends on each of $\mathbf{y}_{bot}$, $\mathbf{y}_{top}$, and $L_x$ in three different ways, and these dependences can be seen when we write the gradient of Nu with respect to $\mathbf{y}_{bot}$, $\mathbf{y}_{top}$, and $L_x$ using the chain rule:
\begin{align}
    \frac{d\mbox{Nu}}{d\mathbf{y}_{bot}} &= \frac{\partial\mbox{Nu}}{\partial \mathbf{y}_{bot}} + \frac{d\mbox{Nu}}{d\mathbf{T}}\frac{d\mathbf{T}}{d\mathbf{y}_{bot}}+ \frac{d\mbox{Nu}}{d\mathbf{T}}\frac{d\mathbf{T}}{d\mathbf{\Psi}}\frac{d\mathbf{\Psi}}{d\mathbf{y}_{bot}}, \label{gradientybot} \\
    \frac{d\mbox{Nu}}{d\mathbf{y}_{top}} &= \frac{\partial\mbox{Nu}}{\partial \mathbf{y}_{top}} + \frac{d\mbox{Nu}}{d\mathbf{T}}\frac{d\mathbf{T}}{d\mathbf{y}_{top}}+ \frac{d\mbox{Nu}}{d\mathbf{T}}\frac{d\mathbf{T}}{d\mathbf{\Psi}}\frac{d\mathbf{\Psi}}{d\mathbf{y}_{top}}, \label{gradientytop} \\
    \frac{d\mbox{Nu}}{dL_x} &= \frac{\partial\mbox{Nu}}{\partial L_x} + \frac{d\mbox{Nu}}{d\mathbf{T}}\frac{d\mathbf{T}}{dL_x}+ \frac{d\mbox{Nu}}{d\mathbf{T}}\frac{d\mathbf{T}}{d\mathbf{\Psi}}\frac{d\mathbf{\Psi}}{dL_x}. \label{gradientLx}
\end{align}
The first terms on the right hand sides of (\ref{gradientybot})--(\ref{gradientLx}) come from the dependences of the differential and integral operators in Nu on $\mathbf{y}_{bot}$, $\mathbf{y}_{top}$, and $L_x$. The second terms are the dependences through $\mathbf{T}$ via operators in the advection-diffusion equation, and the third terms are the dependences through the denominator of $\mathbf{\Psi}$ and then through $\mathbf{T}$ via $\mathbf{\Psi}$ in the advection-diffusion equation.

The optimization algorithm requires the gradient of Nu with respect to $\{\mathbf{c}, \mathbf{B_1},\mathbf{B_2}, A_1, A_2, L_0\}$. These can be written as: 
\begin{align}
    \frac{d\mbox{Nu}}{d\mathbf{c}} &= \frac{d\mbox{Nu}}{d\mathbf{T}}\frac{d\mathbf{T}}{d\mathbf{\Psi}}\frac{d\mathbf{\Psi}}{d\mathbf{c}}, \label{gradientc} \\ 
     \frac{d\mbox{Nu}}{d\mathbf{B}_1} &=  \frac{d\mbox{Nu}}{d\mathbf{y}_{bot}}\frac{d\mathbf{y}_{bot}}{d\mathbf{B}_1} \; ; \;\frac{d\mbox{Nu}}{dA_1} =  \frac{d\mbox{Nu}}{d\mathbf{y}_{bot}}\frac{d\mathbf{y}_{bot}}{dA_1}, \label{gradientybotparams} \\
   \frac{d\mbox{Nu}}{d\mathbf{B}_2} &=  \frac{d\mbox{Nu}}{d\mathbf{y}_{top}}\frac{d\mathbf{y}_{top}}{d\mathbf{B}_2} \; ; \;\frac{d\mbox{Nu}}{dA_2} =  \frac{d\mbox{Nu}}{d\mathbf{y}_{top}}\frac{d\mathbf{y}_{top}}{dA_2}, \label{gradientytopparams} \\
    \frac{d\mbox{Nu}}{dL_0} &=  \frac{d\mbox{Nu}}{dL_x} \frac{dL_x}{dL_0}. \label{gradientL0}
\end{align}
where we have used the left-hand sides of (\ref{gradientybot})--(\ref{gradientLx}) to make 
(\ref{gradientybotparams})--(\ref{gradientL0}) relatively simple.

We can replace the derivatives of $\mathbf{T}$ with respect to $\mathbf{\Psi}$, $\mathbf{y}_{bot}$, $\mathbf{y}_{top}$, and $L_x$ in (\ref{gradientybot})--(\ref{gradientc}) with quantities that are simpler to compute by using the ``adjoint method." 
First we write the discretized advection-diffusion equation (\ref{AdvDiff}) as
\begin{equation}
\mathbf{r}(\mathbf{T},\mathbf{\Psi}, \mathbf{y}_{bot}, \mathbf{y}_{top}, L_x) = 0 \label{Residual}
\end{equation}
with $\mathbf{r}$ (the ``residual") a vector that takes values at each interior grid point. We can think of 
(\ref{Residual}) as an implicit equation for the temperature field $\mathbf{T}$ as a function of $\mathbf{\Psi}$, $\mathbf{y}_{bot}$, $\mathbf{y}_{top}$, and $L_x$. If we apply small perturbations $\Delta\mathbf{\Psi},\dots, \Delta L_x$ to these four variables, we obtain a small perturbation $\Delta\mathbf{T}$ to the temperature field such that the residual remains zero:
\begin{equation}
0 = \Delta \mathbf{r} \approx \frac{d\mathbf{r}}{d\mathbf{T}}\Delta \mathbf{T} +\frac{d\mathbf{r}}{d\mathbf{\Psi}}\Delta\mathbf{\Psi} 
+\frac{d\mathbf{r}}{d\mathbf{y}_{bot}}\Delta\mathbf{y}_{bot} 
+\frac{d\mathbf{r}}{d\mathbf{y}_{top}}\Delta\mathbf{y}_{top}
+ \frac{d\mathbf{r}}{dL_x}\Delta L_x. \label{dr}
\end{equation}
In the limits that the small perturbations tend to zero, we obtain
\begin{equation}
\frac{d\mathbf{T}}{d\mathbf{\Psi}} = -\frac{d\mathbf{r}}{d\mathbf{T}}^{-1}\frac{d\mathbf{r}}{d\mathbf{\Psi}} \; , \; \frac{d\mathbf{T}}{d\mathbf{y}_{bot}} = -\frac{d\mathbf{r}}{d\mathbf{T}}^{-1}\frac{d\mathbf{r}}{d\mathbf{y}_{bot}}  \; , \; 
\frac{d\mathbf{T}}{d\mathbf{y}_{top}} = -\frac{d\mathbf{r}}{d\mathbf{T}}^{-1}\frac{d\mathbf{r}}{d\mathbf{y}_{top}} \; , \; 
\frac{d\mathbf{T}}{dL_x} = -\frac{d\mathbf{r}}{d\mathbf{T}}^{-1}\frac{d\mathbf{r}}{dL_x}.
 \label{dTdPsiLx}
\end{equation}
When these expressions are inserted in (\ref{gradientybot})--(\ref{gradientc}), in each case
we obtain a product 
\begin{align}
\frac{d\mbox{Nu}}{d\mathbf{T}}\frac{d\mathbf{r}}{d\mathbf{T}}^{-1} \equiv \mathbf{\eta}^T
\end{align}
\nn which we have defined as $\mathbf{\eta}^T$, where $\mathbf{\eta}$ is the ``adjoint variable." It is easy to compute $\mathbf{\eta}$ by solving
the ``adjoint equation"
\begin{equation}
    \frac{d\mathbf{r}}{d\mathbf{T}}^T \mathbf{\eta} = \frac{d\mbox{Nu}}{d\mathbf{T}}^T \label{eta}
\end{equation}
with a cost that is essentially the same as solving the advection-diffusion equation. The remaining derivative terms in (\ref{gradientc})--(\ref{gradientL0}) are less expensive to compute since they involve evaluating explicit formulae using the current values of 
$\{\mathbf{T},\mathbf{\Psi}, \mathbf{y}_{bot}, \mathbf{y}_{top}, L_x\}$. Therefore the cost of computing the gradient of Nu is similar to the cost of computing Nu itself; both are dominated by
a large sparse matrix solution (the advection-diffusion equation and its adjoint).

\section*{Acknowledgments}
This research was supported by the NSF-DMS Applied Mathematics program under
award number DMS-2204900.


\end{document}